\documentclass[%
reprint,
amsmath,amssymb,
aps,
]{revtex4-2}

\usepackage{graphicx}
\usepackage{dcolumn}
\usepackage{bm,amsmath}
\usepackage{xcolor}
\usepackage{hyperref}

\begin{document}
\title{Non-Hermitian Nested Hopf-Links and Conjoint Open-Arcs in Synthetic \\ Non-Abelian Gauge Photonic Lattices}
\author{Samit Kumar Gupta}
\email{samitg@iiserbpr.ac.in\,, samit.kumar.gupta@gmail.com}
\affiliation{Department of Physical Sciences, Indian Institute of Science Education and Research Berhampur, Odisha 760003, India}

\begin{abstract}
Non-Hermitian physics enriches the topological attributes of non-Abelian systems that are characterized by noncommutative braid patterns. Non-Abelian braiding of the non-Hermitian bands and anomalous skin mode localization may emerge due to a host of competing physical effects inviting the quest for the generality of their physical origin and the associated new phenomena. Here, we consider a synthetic gauge photonic lattice with multiple competing sources of non-Hermiticity, $\textit{i.e.}$ nearest-neighbor (NN) and next-nearest-neighbor (NNN) hopping mismatches, non-Abelian SU(2) phases, and gain-loss processes. The formation of the distinctive braid patterns and nested Hopf-links is observed, followed by a non-Hermitian topological phase transition at EP and the opening of an imaginary gap beyond. The PBC and OBC eigenspectra and concomitant localization dynamics of the OBC eigenstates show rich physical features compared to their less complex counterparts. This includes the formation of conjoint open-arcs in the OBC spectra giving rise to a completely localized $\textit{pure}$ dipole skin effect without any extended modes and dipole-like skin effect with a few number of extended modes. This work sheds light on some of the key aspects of the synthetic non-Abelian gauge photonic systems in the presence of multiple competing non-Hermitian degrees of freedom that may stimulate further research with applications in robust and fault-tolerant communication channels, computing platforms, and non-Hermitian topological devices.
\end{abstract}
\maketitle
\section{\label{sec:level1} Introduction}
Non-Hermitian physics has provided a novel impetus in the new physics exploration in optics, photonics, condensed matter, and beyond \cite{el2018non,feng2017non,ozdemir2019parity,konotop2016nonlinear, gupta2020parity}. It has captivated attention particularly due to the non-Hermitian topology of exceptional points \cite{ashida2020non,li2023exceptional, ghatak2019new, torres2019perspective, miri2019exceptional}. Several recent activities on this underlying synergy have already revealed a host of intriguing effects and phenomena in the theoretical \cite{bergholtz2021exceptional,ding2022non,leykam2017edge} and experimental \cite{ruter2010observation,peng2016chiral,chen2020revealing} realms. Some of the most prominent effects are topological winding around non-Hermitian singularities \cite{zhong2018winding,midya2018non} and topological signatures of non-Hermitian skin effects \cite{martinez2018non,yao2018edge, kunst2018biorthogonal, jin2019bulk,  okuma2020topological,zhang2022review,zhu2020photonic,okuma2023non,lin2023topological}. Examples may include the relation between topological winding and skin modes \cite{zhang2020correspondence}, critical skin effects \cite{li2020critical, qin2023universal, wang2023scaling}, nonlinearity and interaction-induced skin effects \cite{yuce2021nonlinear,shen2022non}, entanglement phase transition \cite{kawabata2023entanglement}, higher-order skin analogs \cite{zhang2021observation}, and concomitant universality \cite{zhang2022universal}. On the other hand, non-Hermitian topology of exceptional points is inherently related to the non-Abelian energy braiding \cite{wang2021topological,guo2023exceptional,zhang2022non,patil2022measuring,yang2024non,guria2024resolving}, gain-loss-induced braid transition \cite{midya2023gain}, and the knotted topological structures \cite{hu2021knots,cao2024observation}.

A certain class of synthetic gauge photonic systems may prove to be a fertile ground for exploring the non-Abelian dynamics of non-Hermitian systems in new unknown territories. It includes, for example, the complex braiding of non-Hermitian bands in the presence of non-Abelian synthetic gauge fields with nearest-neighbor (NN) interactions \cite{pang2024synthetic} that yields a distinctive signature of simultaneously coexisting left- and right-localized skin localization, and extended modes in stark contrast to one-sided skin localization or extended modes usually observed in conventional nonreciprocal non-Hermitian systems. It is expected that complex wave interaction due to simultaneous consideration of physically different sources of non-Hermiticity, and gain-loss-induced exceptional points, may play an important role in inducing a host of novel complex topological phenomena in the synthetic non-Abelian gauge photonic systems usually absent in the less complex counterparts. However, in the existing literature, a complete picture is still missing. It motivates us to consider this particular point in quest of novel physical insights of such systems in generic configurations to shed new light on the at will control of non-Abelian dynamics and skin localization. In doing so, the following questions are considered. What new features can be thought of to be induced when one considers NNN hoppings in addition to NN interactions with and without onsite gain/loss? What is the role of onsite gain/loss on the braiding pattern formation, the PBC and OBC spectra, and the localization of the OBC eigenstates in presence of competing sources of non-Hermiticity such as NN and NNN interactions and non-Abelian gauge fields? 

On the other hand, a closely related phenomenon in two-band topological systems is the Hopf phases of matter that depict a distinctive delicate topology \cite{pak2024pt,yang2019non,kim2023realization,nakamura2025non}. In general, the Hopf topological phase is unstable in generic non-Hermitian settings. However, PT symmetry has been found to allow homotopy classification of Hopf topological invariants \cite{pak2024pt}. In our non-Abelian two-band non-Hermitian model, new non-Hermitian Hopf phases can be expected to emerge given the increasingly generic nature of the setting. As we will see in our model, formation of two slightly different types of nested Hopf-links have been observed: one when non-Hermiticity is induced only by NNN hopping mismatches and non-Abelian SU (2) phases, and the other when it is induced by three competing sources of non-Hermiticity, namely, hopping mismatches, non-Abelian phases, and onsite gain and loss. The former does not include any EP-mediated sharp topological phase transition, whereas the latter shows a non-Hermitian topological phase transition at EP where one of the interlacing Hopf-link loop structures vanishes. This EP-mediated non-Hermitian topological phase transition of the braided Hopf-links is captured by a topological linking number (see Appendix \ref{AppendixA}) where the EP seems to have induced a topological phase transition from the nontrivial Hopf-link spectral braiding to the trivial unbraiding structure (see Subsection \ref{subsec:level2_nested}).

\begin{figure}
    \centering
    \includegraphics[width=0.99\linewidth]{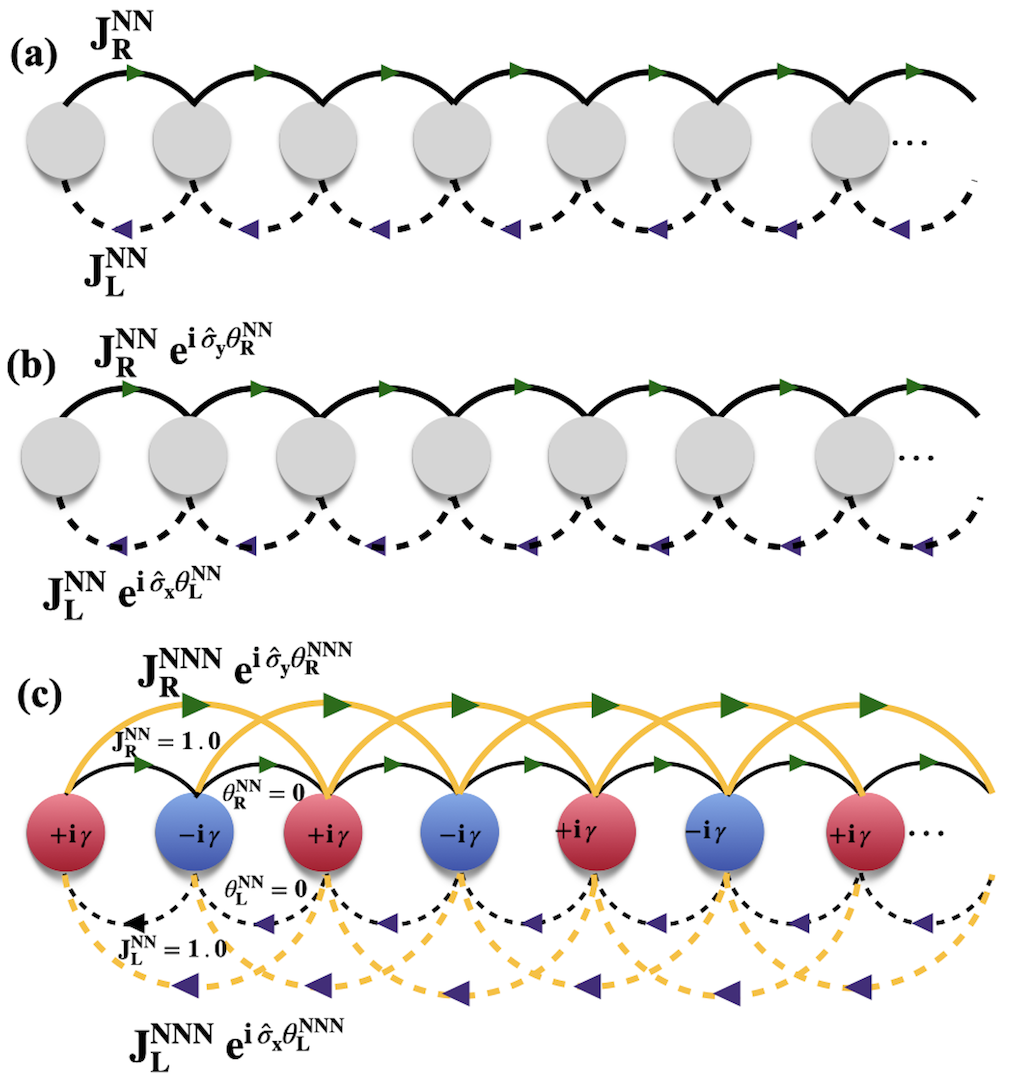}
    \caption{Schematic diagram of the system based on the 1D Hatano-Nelson model with various cases of hopping interactions and non-Hermiticity. (a) Conventional Hatano-Nelson model, (b) Hatano-Nelson model with non-Abelian gauge fields in the NN interactions, (c) the model in our case with NN and NNN hopping interactions, non-Abelian SU (2) phases and gain/loss non-Hermiticity. The solid black lines represent right-ward NN hoppings whereas dashed black lines represent left-ward NN hoppings. The solid yellow lines represent right-ward NNN hoppings whereas dashed yellow lines represent left-ward NNN hoppings. The gray, red, and blue colors represent neutral, gain, and loss sites.}
    \label{fig1}
    \end{figure}

The article is organized as follows: an Introduction is provided in Section I, followed by the “Theoretical Model" in Section II. In Section III, the main results are presented including “Nested Hopf-link formation and EP-mediated topological phase transition" in Subsection III.A, “PBC and OBC eigenspectra, and localization of the OBC eigenstates" in Subsection III.B, and “Conjoint open-arc OBC spectra" in Subsection III.C. Three distinct subcases with respect to the gain/loss distributions in Fig. \ref{fig1}(c) are considered: only lossy end-channel at the edge of the lattice, gain-loss distributions throughout the lattice, and the one with no gain/loss respectively. Finally, conclusion and discussion are provided in Section IV. In addition, related discussion on the topological linking number, effects of system size, and experimental feasibility are given in Appendices A, B, and C respectively. 

\section{\label{sec:level1} Theoretical Model}
In general, simultaneous consideration of NN and NNN hoppings along with onsite gain/loss processes induces increasing complexity into the system. This may, however, reduce the calculation of EP and related non-Hermitian topological dynamics analytically intractable. Such a generic non-Hermitian photonic environment may, nonetheless, lead to exotic complex wave phenomena previously unexplored. Long-range hoppings, in general, can enrich a system’s spectral and localization properties. This is especially true for non-Abelian systems with nonreciprocal hoppings in gain/loss non-Hermitian systems. Apparently, a slightly more simplified model that gives rise to easily tractable closed-form analytic EP to probe the non-Abelian topology of the NH bands via a complex interplay of various coexisting effects such as NN and NNN interactions, non-Abelian SU(2) phases, and onsite gain/loss profiles can be practically designed. It can be argued that despite apparent simplicity of the model, it does not point toward any unrealistic oversimplification, since such settings are physically realizable (see Appendix \ref{AppendixC}). In fact, it is a simplified version of a more generic class of sophisticated synthetic photonic systems allowing us to access the relevant physics across different regimes of EP despite simultaneously considering NNN hoppings, non-Abelian gauge phases, and gain/loss non-Hermiticity. In this case, however, the nonreciprocity in the hopping amplitudes and non-Abelian gauge phases in the NN interaction can be relaxed to simplify the model in analytically accessing EPs. In other words,  $\theta_{L}^{NN}= \theta_{R}^{NN}=0$ and $J_{L}^{NN}=J_{R}^{NN}=1.0$.  
For the sake of completeness, we mention that the real-space tight-binding (TB) Hamiltonian of the system in presence of Abelian phases can be written as:
\begin{equation}
\hat{H}_{R}^{A}
=\sum_{m}{(J_{L}^{NN}\; {\hat{c}^{\dagger}_{m}}e^{i\theta_{L}^{NN}\,} \,{\hat{c}_{m+1}}+J_{R}^{NN} \;{\hat{c}^{\dagger}_{m+1}} e^{i\theta_{R}^{NN}\,} \,{\hat{c}_{m}})}, 
\end{equation}
where, $\theta_{L}^{NN}$, $\theta_{R}^{NN}$ are scalar Abelian phases without any SU(2) spin structure. This returns the conventional Hatano-Nelson (H-N) model for $\theta_{L}^{NN}=\theta_{R}^{NN}=0$. 
On the other hand, the non-Abelian SU(2) phases induce intrinsic two-band spin structures as $\hat\theta^{NN}_{L,R}=\theta^{NN}_{L,R}\;\hat\sigma_{y,x}$ and $\hat\theta^{NNN}_{L,R}=\theta^{NNN}_{L,R}\;\hat\sigma_{y,x}$ in Eq. \ref{eq:na} and Eq. \ref{eq:na_r} , respectively. In the presence of the non-Abelian SU(2) phases in the left-ward or right-ward nonreciprocal NN hoppings, the TB real-space lattice Hamiltonian turns out to be: 
\begin{equation}{\label{eq:na}}
    H^{NA}_{R}=\sum_{m}{(J_{L}^{NN}\; \hat{c}^{\dagger}_{m}}e^{i\theta_{L}^{NN}\,\hat{\sigma}_{y}} \,{\hat{c}_{m+1}}+ J_{R}^{NN} \;\hat{c}^{\dagger}_{m+1} e^{i\theta_{R}^{NN}\,\hat{\sigma}_{x}} \,\hat{c}_{m})
\end{equation}
with its eigenvalues as: 
\begin{align}
E_{\pm}&= A(k) \; \pm \; i\sqrt{(X^2+Y^2)}, \\
X^{2}&=(J_{L}^{NN})^2 \;sin^2{\theta_{L}^{NN}} \,e^{2ik},\\
Y^{2}&=(J_{R}^{NN})^2 \;sin^2{\theta_{R}^{NN}}\,e^{-2ik}.\\
A(k)&=(J_{L}^{NN} \, cos\theta_{L}^{NN}\; e^{ik}+J_{R}^{NN} \, cos\theta_{R}^{NN}\;e^{-ik}).
\end{align}
The condition for the existence of EPs in the quasimomentum reads $(J_{L}^{NN})^2 \;sin^2{\theta_{L}^{NN}} = (J_{R}^{NN})^2\;sin^2{\theta_{R}^{NN}}$ for which the EPs occur at the momenta points $k_{EP}=(\pm \frac{\pi}{4},\pm\frac{3\pi}{4})$ for $k\in(-\pi, \pi)$ as in \cite{pang2024synthetic}. In the more generalized version of the model, the corresponding real-space lattice Hamiltonian of the system with NN and NNN hoppings, non-Abelian SU (2) phases, and onsite gain/loss profiles, in the second-quantized form, can be written as: 
\begin{widetext}{\label{eq:na_r}}
\begin{align}
\hat{H}_R&=\sum_{m}{(J_{L}^{NN}\; \hat{c}^{\dagger}_{m}}e^{i\theta_{L}^{NN}\hat{\sigma}_{y}} \;\hat{c}_{m+1}+J_{R}^{NN} \;\hat{c}^{\dagger}_{m+1} e^{i\theta_{R}^{NN}\hat{\sigma}_{x}} \;\hat{c}_{m})+\sum_{m}(J_{L}^{NNN} \; \hat{c}^{\dagger}_{m}e^{i\theta_{L}^{NNN}\hat{\sigma}_{y}}\; \hat{c}_{m+2}
+J_{R}^{NNN} \;\hat{c}^{\dagger}_{m+2} e^{i\theta_{R}^{NNN}\hat{\sigma}_{x}}\;\hat{c}_{m})\nonumber \\        
& \quad +\sum_{m}{i \;\gamma(-1)^{m} \; \;\hat{c}^{\dagger}_{m} \hat{c}_{m}}
\label{eq:wideeq1}
\end{align}
\end{widetext}


where, the first term under summation on the right-hand side of Eq. \ref{eq:wideeq1} refers to the interaction term due to NN hoppings, the second term under summation refers to the interaction term due to NNN hoppings, and the last one as the gain/loss term.
Its underlying physics can be captured by the following $2\times 2$ k-space effective Hamiltonian with SU(2) non-Abelian gauge fields for the system under periodic boundary condition (PBC), when, for the sake of simplicity, we assume $\theta_{L}^{NN}= \theta_{R}^{NN}=0$ and $J_{L}^{NN}=J_{R}^{NN}=1.0$ to avoid non-Hermiticity originating herein. The non-Abelian gauge fields in NN and NNN hoppings in Eq. \ref{eq:na} and \ref{eq:na_r} induce non-Hermiticity due to directional coupling. The onsite gain/loss distribution basically introduces an alternately gain-loss staggered imaginary gauge potential across the lattice: $V=i\,\gamma\, \times\;\textbf{diag}(1,\,-1,\,1\,,-1\,...(-1)^N).$
\begin{align}
H_{eff}(k)&=(A_{1}+A_{2}) \; \hat{\sigma}_{0}+ i \,\hat{\sigma}_{y}\,J_{L}^{NNN} \; sin\theta_{L}^{NNN}\; e^{2ik}\nonumber \\
& +i \, \hat{\sigma}_{x} \,J_{R}^{NNN} \;sin\theta_{R}^{NNN} \; e^{-2ik}+i\, \gamma\hat{\sigma}_{z},
\label{eq:model_eq}
\end{align}
\begin{align}
A_{1}&=(J_{L}^{NN}e^{ik}+J_{R}^{NN}e^{-ik}), \nonumber \\
A_{2}&=(J_{L}^{NNN} \; cos\theta_{L}^{NNN}\; e^{2ik}+J_{R}^{NNN} \; cos\theta_{R}^{NNN}\; e^{-2ik}), \nonumber
\end{align}
where, $\hat{c}_{m}^{\dagger}/ \hat{c}_{m}$ are the creation/annihilation operators, respectively. Here, $\hat{\sigma}_{x}=[0,\,-1;\;1,\,0], \hat{\sigma}_{y}=[0,\,-i;\;i,\,0], \hat{\sigma}_{z}=[1,\,0;\;0,\,-1]$ are Pauli matrices and $\hat{\sigma}_{0}=\hat{I}=[1,\,0;\;0,\,1],$ $J_{L}^{NNN}$ and $J_{R}^{NNN}$ are left- and right-ward NNN hoppings, $\theta_{L}^{NN}$ and $\theta_{R}^{NN}$ are left- and right-ward NN non-Abelian SU (2)phases, $\theta_{L}^{NNN}$, and $\theta_{R}^{NNN}$ are left- and right-ward NNN non-Abelian phases, where N represents the total number of sites, and $\gamma$ refers to the onsite gain/loss. 
Its eigenvalues are found to be:   
\begin{align}  \label{eq:model_eq_eig}
E_{\pm}&=(A_{1}+A_{2}) \, \pm \, i \sqrt{(X_1^2+X_2^2+\gamma^2)},  \\
X_1^2&=(J_{L}^{NNN})^2 \;sin^2{\theta_{L}^{NNN}} \, e^{4ik}, \\
X_2^2&=(J_{R}^{NNN})^2 \;sin^2{\theta_{R}^{NNN}} \;e^{-4ik}.
\end{align}  
The Hamiltonian in Eq. \ref{eq:model_eq} is PT symmetric, respecting $\mathcal{PT}H_{eff}(k)(\mathcal{PT})^{-1}=H_{eff}(k)$ under reciprocal limits of the NN and NNN couplings $J_L^{NN}=J_{R}^{NN},\theta_L^{NN}=\theta_{R}^{NN}, J_L^{NNN}=J_{R}^{NNN}, \theta_L^{NNN}=\theta_{R}^{NNN}$. It is straightforward to show that in the absence of gain/loss non-Hermiticity, EP occurs at an increasingly more number of momenta points at $k_{EP}=({\pm\frac{\pi}{8},\pm\frac{3\pi}{8},\pm\frac{5\pi}{8},\pm\frac{7\pi}{8}})$ for $k\in(-\pi,\pi)$ which is a direct consequence of the NNN interactions via hopping amplitudes and non-Abelian SU(2) phases. Here, the condition for EP to occur is $(J_{L}^{NNN})^2 \;sin^2{\theta_{L}^{NNN}} = (J_{R}^{NNN})^2 \;sin^2{\theta_{R}^{NNN}}$ which is similar to the case where only the NN interaction is present. It allows us to obtain the following closed-form expression of EP entirely dictated by the NNN hoppings and non-Abelian SU(2) phases: 
\begin{equation}
\gamma_{EP}^{NNN}=\gamma_{EP}=\sqrt{\; 2 \;J_{L}^{NNN} \; J_{R}^{NNN}\; sin{\theta_{L}^{NNN}} \; sin{\theta_{R}^{NNN}}}.
\label{eq:EP_NNN}
\end{equation}
We may note that in the situation where $\theta_{L}^{NN} \neq\theta_{R}^{NN}\ne 0$ and $J_{L}^{NN}= J_{R}^{NN} = 1.0$, it comparatively yields a much more complex eigenstructure where the analytical finding of closed-form EP becomes quite difficult. In general, such a closed-form analytic EP makes it possible to probe the distinctive non-Hermitian EP phase transition and the concomitant spectral braid formation dynamics across it. As we will see in the course of this paper, this analytical prediction of EP-mediated topological phase transition is also numerically demonstrated via a topological linking number that undergoes a phase transition from a nontrivial to a trivial topological phase. It could be worth mentioning here that this particular form of EP is also derivable in the gain/loss-mediated H-N model with only NN hoppings and non-Abelian gauge phases, with the obvious difference being that hopping amplitudes and non-Abelian gauge phases are replaced by their NN analogs, as given below:   
\begin{align}
E_{\pm}&=(A_{1}+A_{2}) \; \pm \; i\sqrt{(Y_1^2+Y_2^2+\gamma^2)}, \\
Y_1^2&=(J_{L}^{NN})^2 \;sin^2{\theta_{L}^{NN}} \;e^{2ik}, \\
Y_2^2&=(J_{R}^{NN})^2 \;sin^2{\theta_{R}^{NN}} \;e^{-2ik}, \\
\gamma_{EP}^{NN}&=\sqrt{\; 2 \;J_{L}^{NN} \; J_{R}^{NN}\; sin{\theta_{L}^{NN}} \; sin{\theta_{R}^{NN}}}.
\end{align}

It is remarkable to see that despite the presence of NNN hoppings, its eigenstructure is pretty much equivalent to it under the presumption that the NN hoppings are reciprocal, \textit{i.e.} $J_L^{NN}=J_R^{NN}$. However, their eigenproperties are governed in a very similar way. This provides us with a direct structural proximity of these two systems with similar (if not exactly the same) physical behaviors even though they appear to be quite different.    
On the other hand, if $\theta_{L}^{NN}  \neq 0, \;\theta_{R}^{NN}\ne 0$, it leads to a Hamiltonian $H_{Eff}^{Gen}(k)$ with structurally complicated eigenstructure for which accessing EPs becomes difficult. In this case, the effective Hamiltonian of the system can be found as:
\begin{equation}
H_{eff}^{Gen}(k)=(A_{1}+A_{2})\; \hat{\sigma}_{0}+H_{x}+H_{y}+H_{NH}, \\
\end{equation}
\begin{align}
H_{x}&=i \;\hat{\sigma}_{x}(\,J_{R}^{NN} \;sin\theta_{R}^{NN} e^{-ik}\;+J_{R}^{NNN} \;sin\theta_{R}^{NNN} \; e^{-2ik}), \nonumber \\
H_{y}&=i \;\hat{\sigma}_{y}(\,J_{L}^{NN} \; sin\theta_{L}^{NN}\; e^{ik}+J_{L}^{NNN} \; sin\theta_{L}^{NNN}\; e^{2ik}),\nonumber \\
H_{NH}&=i\; \gamma \,\hat{\sigma}_{z} \nonumber.
\end{align}
The eigenvalues of this system can be written as: 
\begin{equation}
E^{Gen}_{\pm}=A_1(k)+A_2(k)\pm i\,\sqrt{F^2(k)+\gamma^2}
\label{eq:EP_gen}
\end{equation}
with each of the terms given below: 
\begin{widetext}
\begin{align}
-F^2(k) &=F_{1}(k)+F_{2}(k),\\
F_{1}(k) &=-J_{L1}^2e^{2ik}-J_{L2}^2e^{4ik}-J_{R1}^2e^{-2ik}-J_{R2}^2 e^{-4ik}, \\
F_{2}(k) &=-2J_{L1} J_{L2}\, e^{3ik}+iJ_{L1}J_{R1}+J_{L1} J_{R1}\, e^{-ik} 
+iJ_{L2} J_{R1}\, e^{ik}+ 
iJ_{L2}J_{R2}-iJ_{L1}J_{R1}-iJ_{L2}J_{R1}e^{ik}-2J_{R1} J_{R2}\, e^{-3ik}\nonumber \\ 
         & \quad -iJ_{L1} J_{R2}e^{-ik} -iJ_{L2} J_{R2},
\label{eq:wideeq2}
\end{align}
\end{widetext}
\begin{figure*}
    \centering
    \includegraphics[width=0.96\linewidth]{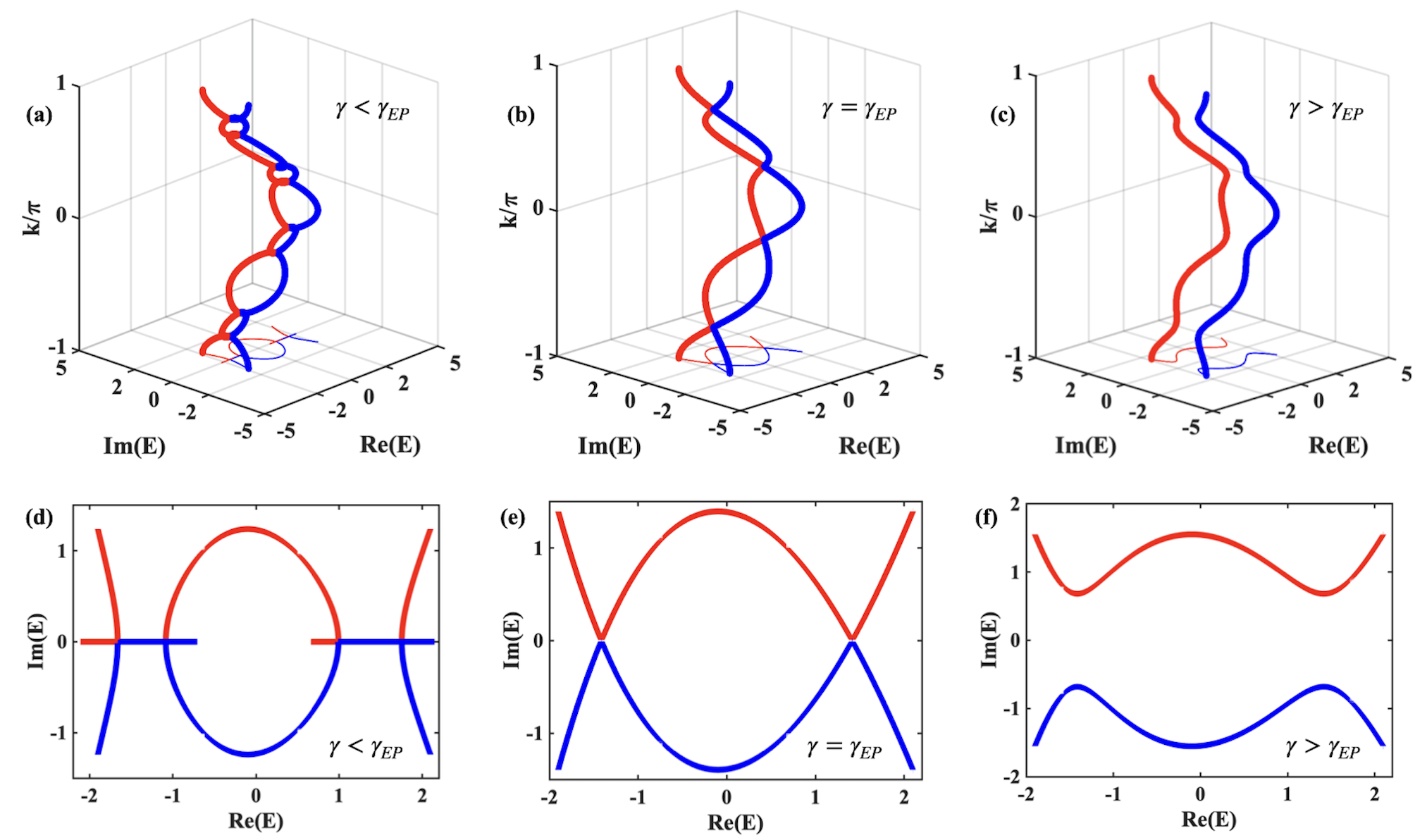}
    \caption{Formation of the intertwining nested Hopf-links and the non-Hermitian topological phase transition. The upper row panels show the evolution trajectories of the particle in the complex plane $Re(E)-Im(E)$ where $E=E_{PBC}=E_{\pm}\in \mathbb{C}$ as in Eq. \ref{eq:model_eq_eig} as the quasimomentum $k$ is varied between $-\pi$ and $+\pi$ whereas the lower row panels show the corresponding projection onto the complex plane. (a,d)\, $\gamma=0.75, (b,e)\, \gamma=0.9875, (c,f)\, \gamma=1.2. $ \, Other parameters:\, $J_{L}^{NN}=J_{R}^{NN}=1.0,\, J_{L}^{NNN}=J_{R}^{NNN}=0.7,\, \theta_{L}^{NN}=\theta_{R}^{NN}=0.0, \, \theta_{L}^{NNN}=\theta_{R}^{NNN}=-1.5$. }
    \label{fig2}
\end{figure*}
where, $J_{L1}=J_{L}^{NN}\,sin\theta_{L}^{NN},\; J_{R1}=J_{R}^{NN}\,sin\theta_{R}^{NN},\;J_{L2}=J_{L}^{NNN}\,sin\theta_{L}^{NNN},J_{R2}=J_{L}^{NNN}\,sin\theta_{R}^{NNN}.$ It is quite understandable from Eq. \ref{eq:EP_gen} that obtaining an analytical expression of EP becomes a cumbersome task unless some suitable numerical approaches are employed. Therefore, for the sake of simplicity, we refrain from considering it further. 
\begin{figure}
    \centering
    \includegraphics[width=0.96\linewidth]{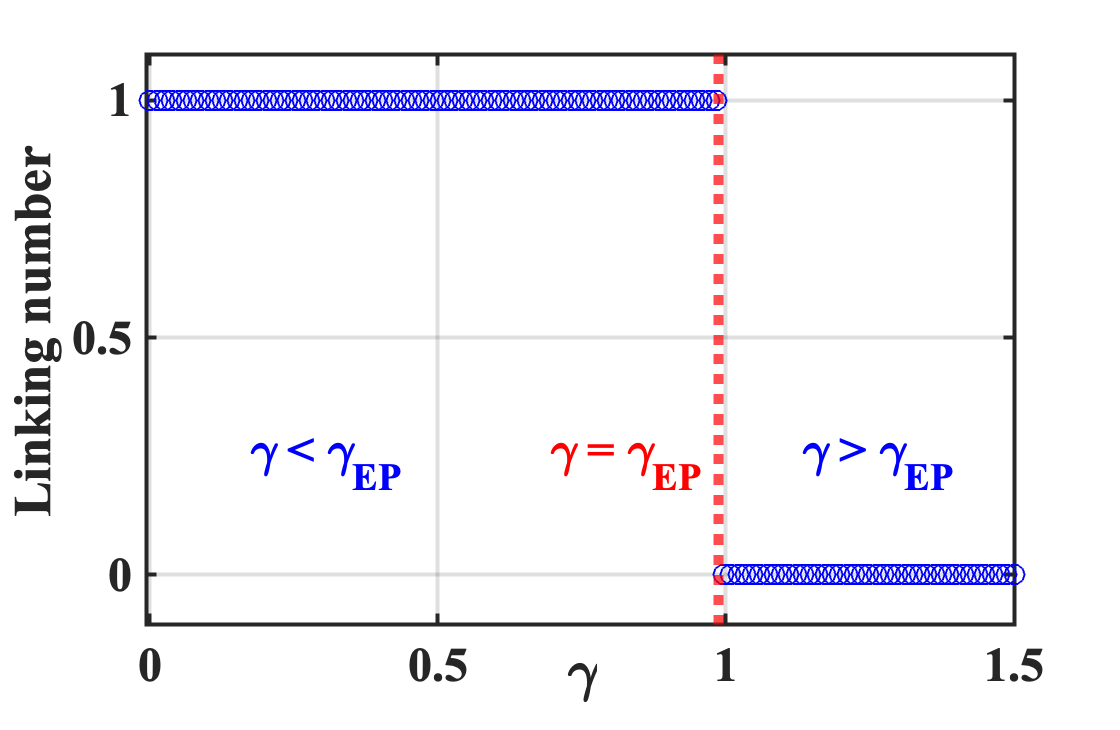}
    \caption{The topological linking number $\mathrm{L}$ (see Appendix \ref{AppendixA}) plotted against the gain-loss non-Hermitian parameter $\gamma$. The red dotted line corresponds to the critical value of $\gamma$ at EP, \textit{i.e.} $\gamma_{EP}=0.9875$ in Eq. \ref{eq:EP_NNN} of the system as in Fig. \ref{fig1}. Other parameters:\, $J_{L}^{NN}=J_{R}^{NN}=1.0,\, J_{L}^{NNN}=J_{R}^{NNN}=0.7,\, \theta_{L}^{NN}=\theta_{R}^{NN}=0.0, \, \theta_{L}^{NNN}=\theta_{R}^{NNN}=-1.5$. The critical value of $\gamma$ at EP $\gamma=\gamma_{EP}$ marks the critical point at which the linking number undergoes a sharp topological phase transition from 1 to 0.}
    \label{fig3}
\end{figure}
    
\section{\label{sec:level1} Main results}
The main results of this work include the EP-mediated non-Hermitian topological phase transition and intertwining nested Hopf-links, localization dynamics of the OBC eigenmodes and skin localization, and revelation of the conjoint open-arc OBC spectra. 

\subsection{\label{subsec:level2_nested} Nested Hopf-link formation and EP-mediated topological phase transition}
So far, we have seen that our model based on the non-Hermitian k-space effective Hamiltonian in Eq. \ref{eq:model_eq} yields the complex eigenenergies and associated EP in Eq. \ref{eq:model_eq_eig} and Eq. \ref{eq:EP_NNN}, respectively. As shown in Fig. \ref{fig2}, variation of gain/loss non-Hermiticity $\gamma$ induces distinctive braiding patterns for which the onsite gain-loss non-Hermiticity bears a direct controllable effect via a EP-mediated topological phase transition. Below EP in the PT unbroken phase $\gamma<\gamma_{EP}$, the evolution trajectories of the particle as $k$ is varied in the Brillouin zone $k\in (-\pi,+\pi)$ in the complex energy plane reveal an intertwining nested Hopf-link-type structure with two interlacing closed loops (Fig. \ref{fig2} (a),(d)). As the system passes through the EP, one of these intertwining Hopf-link loops disappears in the vicinity of $\gamma=\gamma_{EP}$ (Fig. \ref{fig2} (b),(e)), above which an imaginary gap opens with the two bands remaining separated and unbraided (Fig. \ref{fig2} (c),(f)) for $\gamma>\gamma_{EP}$. At EP, the Hopf-link braid structure deforms sharply separating the spectral Hopf-link braiding phase from the non-braiding phase. In order to characterize the topological nature of the EP-mediated phase transition, we introduce a linking number, which is a topological invariant that measures the spectral linking between two spectral trajectories (see Appendix \ref{AppendixA}). This is evident from the fact that the linking number $L$ undergoes a sharp topological phase transition at $\gamma=\gamma_{EP}$ from a topologically nontrivial ($L=1$) to trivial ($L=0$) phases (see Fig. \ref{fig3}). This means that the two spectral trajectories of the complex energy bands are linked for $L=1$ and unlinked for $L=0$. On a physical point of view, this stems from the square-root structures of the complex energy bands giving rise to two coupled Riemann sheets. In the nontrivial topological phase, the sheets remain connected through branch-point structures, leading to linked spectral structures in the embedded $Re(E)-Im(E)$ plane. As the  gain-loss parameter $\gamma$ approaches the EP, the branch connectivity collapses and the two spectral trajectories become disconnected, leading to the topological phase transition of the Hopf-link structure.

We see that the intertwining nested Hopf-link-type structure is also present without gain/loss non-Hermiticity albeit with slightly distinct features. For example, in the presence of gain/loss non-Hermiticity varying $\gamma$ witnesses a EP-mediated topological phase transition where the nested Hopf-link-type structure deforms into a collection of two interlacing non-Hermitian bands crossing each other where one of the intertwining Hopf-links disappears. In the absence of gain/loss non-Hermiticity, the intertwining nested Hopf-link-type structure simply deforms into two separated bands without any disappearing intertwining loops. In some sense, this is equivalent to loss of nontrivial spectral braiding of the non-Hermitian bands. Hence, gain/loss non-Hermiticity seems to lead to the trivialization of the braiding of the non-Hermitian bands in the broken PT phase beyond EP in the presence of non-Abelian synthetic gauge fields.  

\subsection{PBC and OBC eigenspectra, and localization of the OBC eigenstates}
First, we consider the case where there is only one lossy site at the end of the lattice $n=N$ essentially rendering it to a non-Hermitian edge-impurity. For the sake of simplicity, we begin by considering when the NNN hoppings are also reciprocal, \textit{i.e.} $J_{L}^{NNN} = J_{R}^{NNN}$ in addition to $J_{L}^{NN}= J_{R}^{NN} = 1.0$. In this case, the sources of non-Hermiticity is solely due to the lossy end-site. For $\gamma=0.2$, the excitation of the left- or right-localized gain-loss-induced defect \textit{edge} modes occurs leading to the simultaneous excitation of the few-number left- and right-localized edge states. There is an additional key distinctive feature that appears here whose origin is purely induced by lack of gain/loss non-Hermiticity. In the absence of gain/loss non-Hermiticity, $\textit{i.e.}$ when $\gamma=0$, in the absolute reciprocal limit (\textit{i.e.} non-Hermitian features due to hopping mismatch and non-Abelian phases are turned off) all the extended modes are uniformly distributed across the lattice leading to complete delocalization. As $\gamma\rightarrow 0$, these left- and right-localized defect \textit{edge} modes disappear. Hence, it can be inferred that the origin of such left- and right-localized defect \textit{edge} modes is purely induced by gain/loss non-Hermiticity. We will see in the next subsection that it is possible to have a purely left- and -right skin localization without any delocalized extended modes. 

\begin{figure}
    \centering
    \includegraphics[width=0.998\linewidth]{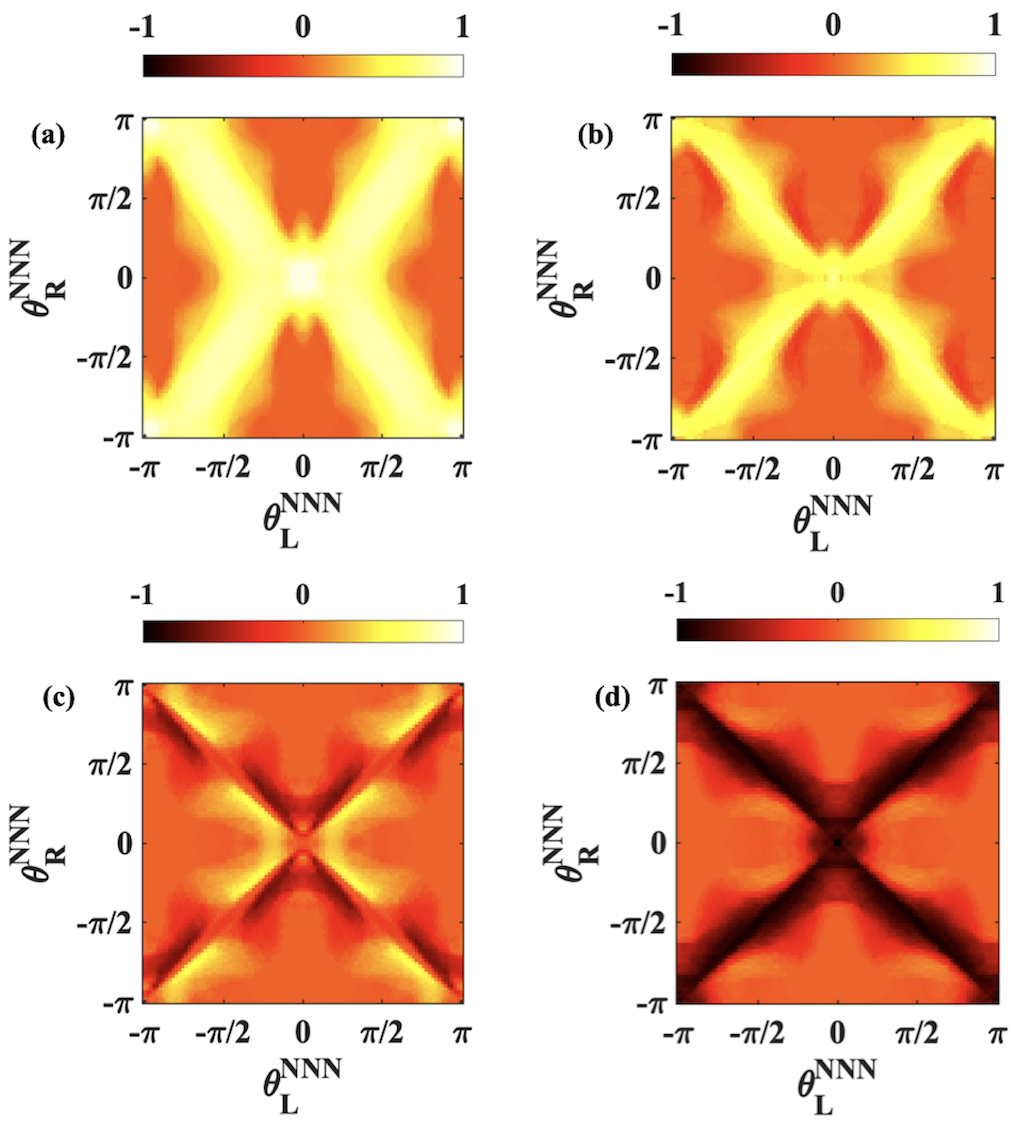}
    \caption{Population contrast $\eta(\theta_{L}^{NNN},\, \theta_{R}^{NNN})$ of the left- ($n_{L}$) and right-localized ($n_{R}$) skin modes, and delocalized extended modes ($n_{E}$) in the $\theta_{L}^{NNN}-\theta_{R}^{NNN}$ plane for the case of lossy end-site non-Hermiticity or single-site non-Hermitian edge-impurity showing unique distribution patterns in different regimes of NNN hopping amplitudes. (a) $(J_{L}^{NNN},J_{R}^{NNN})=(0.7,0.4)$ (b) $(J_{L}^{NNN},J_{R}^{NNN})=(0.7,0.6),$ (c) $(J_{L}^{NNN},J_{R}^{NNN})=(0.7,0.7),$ (d) $(J_{L}^{NNN},J_{R}^{NNN})=(0.6,0.7).$  The value of gain/loss parameter is $\gamma=0.8$, and $J_L^{NN}=J_R^{NN}=1.0$.}
    \label{fig4}
\end{figure}

To shed more light on the localization profiles of the three types of modes, the population contrast $\eta(\theta_{L}^{NNN},\theta_{R}^{NNN})$ is plotted in the parameter plane $\theta_{L}^{NNN}-\theta_{R}^{NNN}$ in Fig. \ref{fig4}. It can be defined as follows: $\eta(\theta_{L}^{NNN},\,\theta_{R}^{NNN})=\frac{(n_L-n_R)}{(n_L+n_R+n_E)}$, where $ n_{L}, n_{R},$ and $n_{E}$ represent the number of left-, right-localized skin modes, and delocalized extended modes with total number of modes $n=n_L+n_R+n_E$. One can find that for $J_L^{NNN}> J_R^{NNN}$ the population contrast distribution yields regimes where either $n_L\approx n_R$ or it is predominantly occupied by left-localized skin modes (Fig. \ref{fig4}(a),(b)). On the other hand, for $J_L^{NNN}< J_R^{NNN}$ the population contrast distribution yields regimes where either $n_L\approx n_R$ or it is predominantly occupied by right-localized skin modes (Fig. \ref{fig4}(d). At the NNN reciprocal limit when $J_L^{NNN}=J_R^{NNN}$ the population contrast distribution yields regimes where either $n_L\approx n_R$ or it is simultaneosuly occupied by the left- and right-localized skin modes (Fig. \ref{fig4}(c). We note that $\eta=0$ can qualitatively refer to two distinct scenarios where either $n_L=n_R$ or $n_E\gg n_L, n_E\gg n_R$. At this stage, a nonzero $\gamma$ does not induce any appreciable alteration in the distributions of $\eta$. On the other hand, a nonzero $\gamma$ induces gain/loss-induced excitation of a \textit{few-number} of the left- and right-localized defect edge modes. In some cases, it appears to aid in the simultaneous recovery of the left- and right-localized edge modes from the collection of extended modes when the source of non-Hermiticity is solely determined by gain/loss distribution. Since this number is quite low, no significant change in their distribution is prominently visible in the population contrast plot in the parameter plane of the non-Abelian gauge phases $\theta_{L}^{NNN}-\theta_{R}^{NNN}$.  

\begin{figure}
    \centering
    \includegraphics[width=0.998\linewidth]{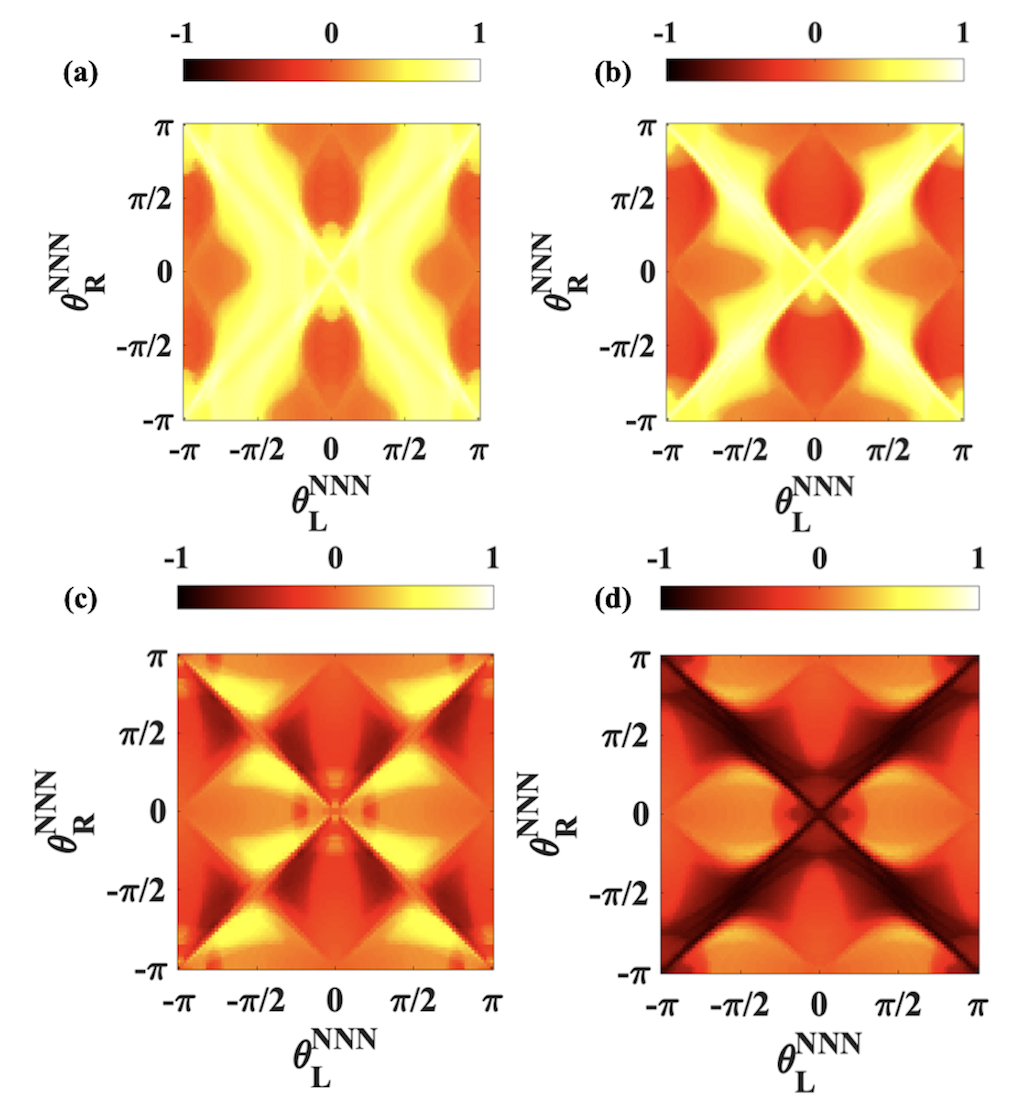}
    \caption{Population contrast $\eta(\theta_{L}^{NNN},\, \theta_{R}^{NNN})$ of the left- ($n_{L}$) and right-localized ($n_{R}$) skin modes, and delocalized extended modes ($n_{E}$) in the $\theta_{L}^{NNN}-\theta_{R}^{NNN}$ plane for the case of alternate gain-loss distributions showing unique distribution patterns in different regimes of NNN hopping amplitudes. (a) $(J_{L}^{NNN},J_{R}^{NNN})=(0.7,0.4)$ (b) $(J_{L}^{NNN},J_{R}^{NNN})=(0.7,0.6),$ (c) $(J_{L}^{NNN},J_{R}^{NNN})=(0.7,0.7),$ (d) $(J_{L}^{NNN},J_{R}^{NNN})=(0.6,0.7).$  The value of gain/loss parameter is $\gamma=0.8$, and $J_L^{NN}=J_R^{NN}=1.0$.}
    \label{fig5}
\end{figure}

For the same set of parameters as in Fig. \ref{fig4}, we have plotted the population contrast in Fig. \ref{fig5} for the case where the gain-loss is distributed alternately across the whole lattice. One can find that for $J_L^{NNN}> J_R^{NNN}$ the population contrast distribution yields regimes where either $n_L\approx n_R$ or it is predominantly occupied by left-localized skin modes (Fig. \ref{fig5}(a),(b)). On the other hand, for $J_L^{NNN}< J_R^{NNN}$ the population contrast distribution yields regimes where either $n_L\approx n_R$ or it is predominantly occupied by right-localized skin modes (Fig. \ref{fig5}(d) by two intersecting perfectly straight diagonal lines. At the NNN reciprocal limit when $J_L^{NNN}=J_R^{NNN}$ the population contrast distribution yields regimes where either $n_L\approx n_R$ or it is simultaneosuly occupied by the left- and right-localized skin modes (Fig. \ref{fig5}(c). Here also, $\eta=0$ may refer to qualitatively two distinct scenarios where either $n_L=n_R$ or $n_E\gg n_L, n_E\gg n_R$. We find that for the no gain-loss case with $\gamma=0$ and for the same set of parameter values, it yields the same distribution patterns of the population contrast as in Fig. \ref{fig4}.

\subsection{Conjoint open-arc OBC spectra}
We first consider the lossy end-site non-Hermiticity case. We have found that the model yields an interesting scenario in which a pair of open-arc-like conjoint OBC spectra appear in the complex phase diagram with \textit{pure} dipole skin effect of left- and right-localized skin modes without any extended modes. This occurs under a specific set of parameter values: $J_{L}^{NN}=J_{R}^{NN}=1.0$, $J_{L}^{NNN}=J_{R}^{NNN}=0.85$, $\theta_{L}^{NN}=\theta_{R}^{NN}=0.0$, $\theta_{L}^{NNN}=-1.8,\, \theta_{R}^{NNN}=-1.0$ and $\gamma=0.2$. In this case, the only sources of non-Hermiticity are due to the imbalanced non-Abelian SU(2) phases and lossy end-site non-Hermiticity $\gamma$. Keeping $\theta_{R}^{NNN}=-1.0$ as $\theta_{L}^{NNN}$ is varied within the range $\theta_{L}^{NNN}\in (-1.0,2.2)$, it first gives rise to a single open-arc OBC structure (Fig. \ref{fig5}(b2)) that eventually leads to the formation of the conjoint open-arc OBC spectra (Fig. \ref{fig5}(e2)). Correspondingly, it still gives rise to the simultaneous left- and right-ward skin localization, and delocalized extended modes (most prominently visible in Fig. \ref{fig5}(a),(b),(c),(d)). However, all complex eigenvalues in the two distinct OBC open-arc-like spectra correspond only to the left- and right-localized skin modes (Fig. \ref{fig5}(e)). This can be termed as \textit{pure} open-arc structure that leads to a \textit{pure} dipole skin effect where all the eigenmodes are left- and right-skin localized without any extended mode delocalization. This perhaps raises the potential question whether there could be some sort of underlying connection between conjoint OBC spectra and a pure dipole skin effect in such systems. This is in stark contrast to some situations where such simultaneous left- and right-localized skin modes, and delocalized extended modes occur due to a single \textit{hybrid} open-arc OBC spectra with all three types of eigenvalues lying on the same open-arc structure (Fig. \ref{fig5}(a),(b),(c),(d)).
    
\begin{figure*}
    \centering
    \includegraphics[width=0.99\linewidth]{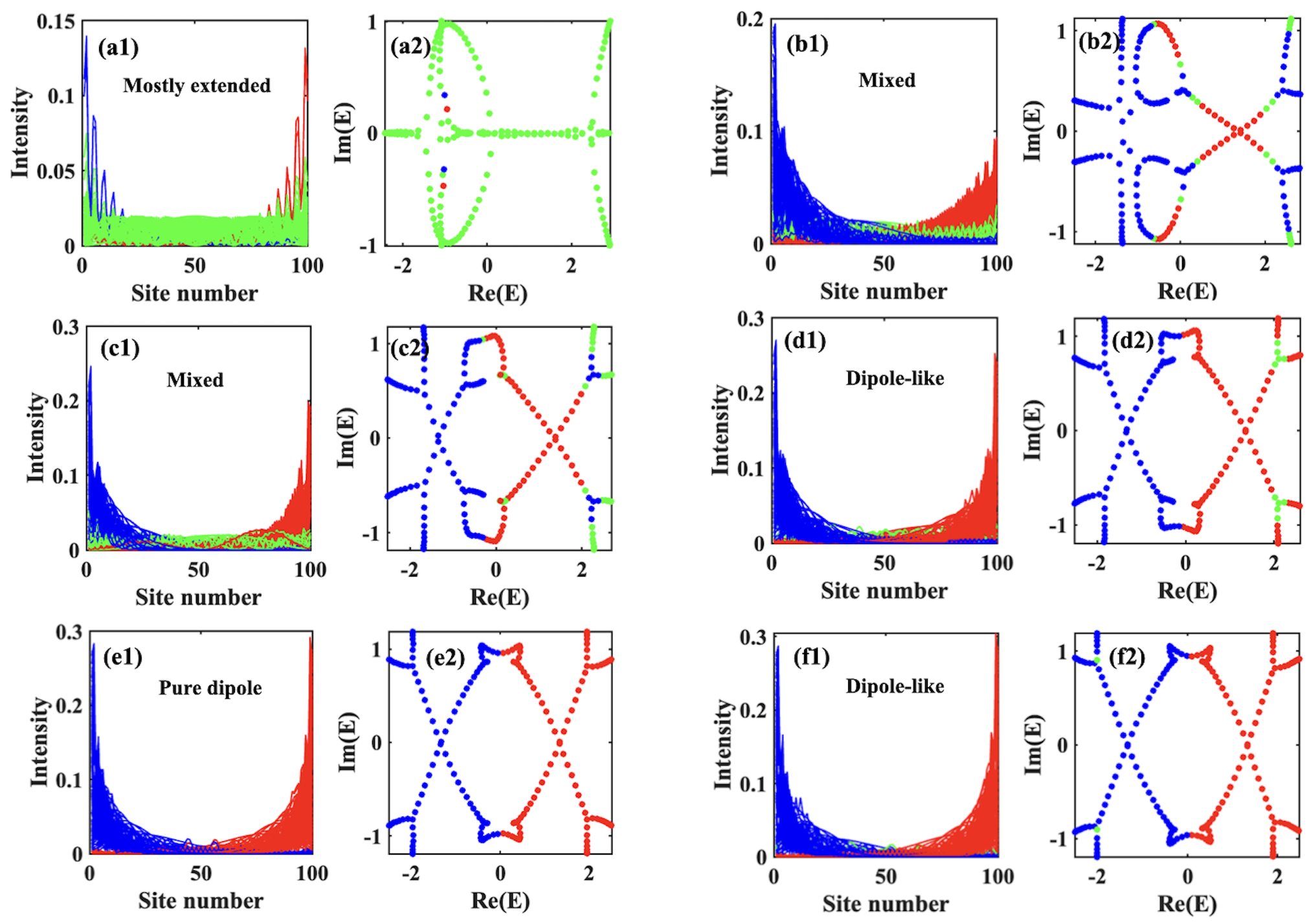} 
    \caption{Different stages of the formation of the conjoint open-arc OBC spectra in the case of lossy end-site non-Hermiticity. (a1,a2) $\theta_{L}^{NNN}=-1.0,$ (b1,b2) $\theta_{L}^{NNN}=-1.4,$ (c1,c2) $\theta_{L}^{NNN}=-1.8,$ (d1,d2) $\theta_{L}^{NNN}=-2.0,$ (e1,e2) $\theta_{L}^{NNN}=-2.15,$ (f1,f2) $\theta_{L}^{NNN}=-2.2.$ Other parameters: $J_{L}^{NN}=J_{R}^{NN}=1.0$, $J_{L}^{NNN}=J_{R}^{NNN}=0.85$, $\theta_{L}^{NN}=\theta_{R}^{NN}=0.0$,\, $\theta_{R}^{NNN}=-1.0$ and $\gamma=0.2$. In all these panels, the subfigures (a1,b1,c1,d1,e1,f1) depict the intensity distributions of the OBC eignestates across the lattice sites $N=100$ (left-localized modes: blue curves, right-localized modes: red curves, extended modes: green curves) and the subfigures (a2,b2,c2,d2,e2,f2) show the corresponding OBC (left-localized modes: blue dots, right-localized modes: red dots, extended modes: green dots in the complex energy plane $Re(E)-Im(E)$ where $E=E_{OBC}\in \mathbb{C}$. Note that due to lack of translational symmetry of this system PBC spectrum is not shown here.}
    \label{fig6} 
\end{figure*}

\begin{figure*}
    \centering
    \includegraphics[width=0.99\linewidth]{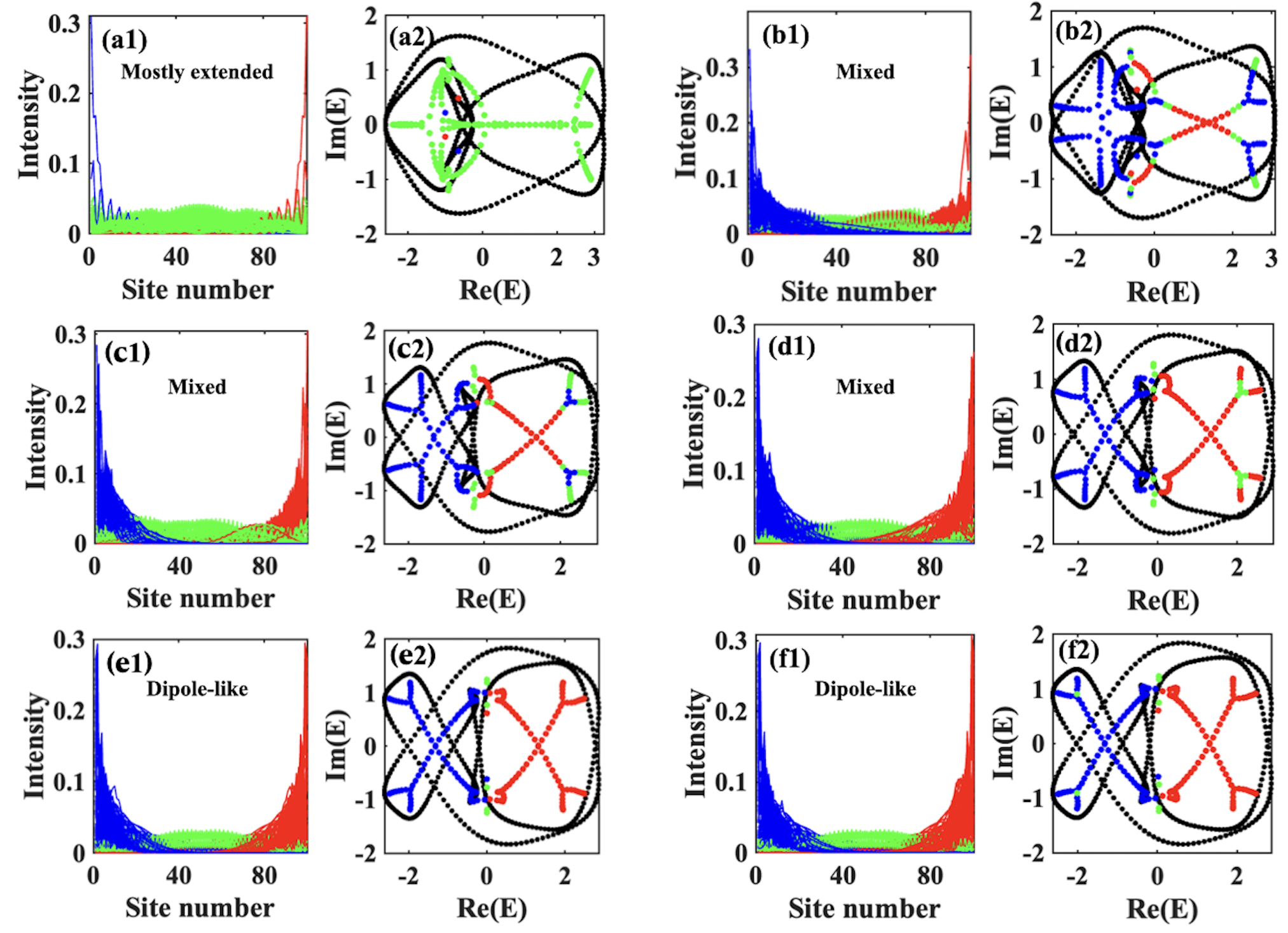} 
    \caption{Different stages of the formation of the conjoint open-arc OBC spectra for the case of alternately distributed gain-loss distributions across the lattice. (a1,a2) $\theta_{L}^{NNN}=-1.0,$ (b1,b2) $\theta_{L}^{NNN}=-1.4,$ (c1,c2) $\theta_{L}^{NNN}=-1.8,$ (d1,d2) $\theta_{L}^{NNN}=-2.0,$ (e1,e2) $\theta_{L}^{NNN}=-2.15,$ (f1,f2) $\theta_{L}^{NNN}=-2.2.$ Other parameters: $J_{L}^{NN}=J_{R}^{NN}=1.0$, $J_{L}^{NNN}=J_{R}^{NNN}=0.85$, $\theta_{L}^{NN}=\theta_{R}^{NN}=0.0$,\, $\theta_{R}^{NNN}=-1.0$ and $\gamma=0.2$. In all these panels, the subfigures (a1,b1,c1,d1,e1,f1) depict the intensity distributions of the OBC eigenstates across the lattice sites $N=100$ (left-localized modes: blue curves, right-localized modes: red curves, extended modes: green curves) and the subfigures (a2,b2,c2,d2,e2,f2) show the corresponding OBC (left-localized modes: blue dots, right-localized modes: red dots, extended modes: green dots and PBC (black dots) eigenspectra in the complex energy plane $Re(E)-Im(E)$ where $E=E_{\pm}\in \mathbb{C}$ as in Eq. \ref{eq:model_eq_eig} for the black dots and $E=E_{OBC}\in \mathbb{C}$ for blue, red, and green dots.}
    \label{fig7} 
\end{figure*}

\begin{figure*}
    \centering
    \includegraphics[width=0.99\linewidth]{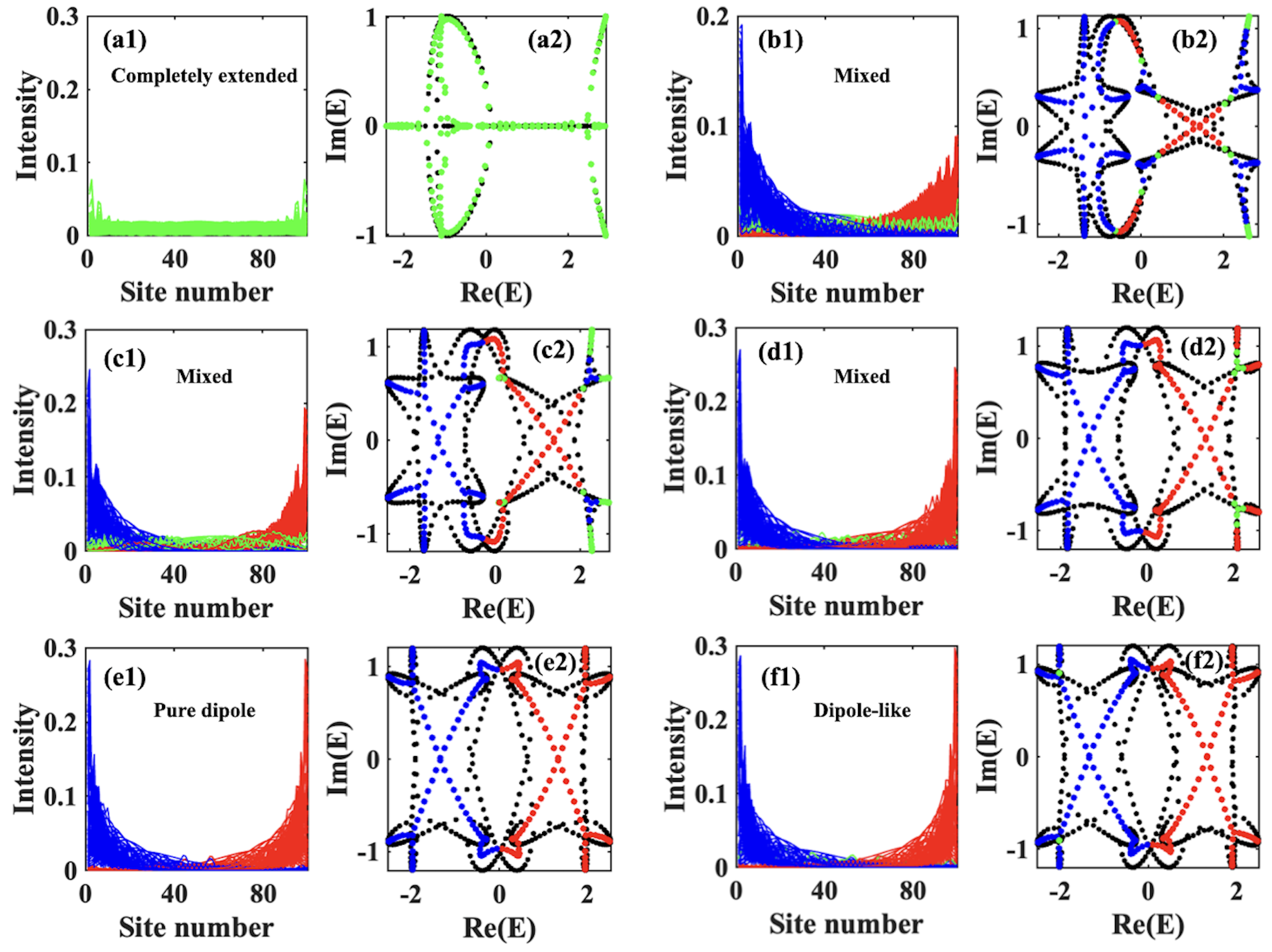} 
    \caption{Different stages of the formation of the conjoint open-arc OBC spectra for the case of no gain-loss distributions across the lattice, \textit{i.e.} $\gamma =0$. (a1,a2) $\theta_{L}^{NNN}=-1.0,$ (b1,b2) $\theta_{L}^{NNN}=-1.4,$ (c1,c2) $\theta_{L}^{NNN}=-1.8,$ (d1,d2) $\theta_{L}^{NNN}=-2.0,$ (e1,e2) $\theta_{L}^{NNN}=-2.15,$ (f1,f2) $\theta_{L}^{NNN}=-2.2.$ Other parameters: $J_{L}^{NN}=J_{R}^{NN}=1.0$, $J_{L}^{NNN}=J_{R}^{NNN}=0.85$, $\theta_{L}^{NN}=\theta_{R}^{NN}=0.0$,\, $\theta_{R}^{NNN}=-1.0$ and $\gamma=0.2$. In all these panels, the subfigures (a1,b1,c1,d1,e1,f1) depict the intensity distributions of the OBC eigenstates across the lattice sites $N=100$ (left-localized modes: blue curves, right-localized modes: red curves, extended modes: green curves) and the subfigures (a2,b2,c2,d2,e2,f2) show the corresponding OBC (left-localized modes: blue dots, right-localized modes: red dots, extended modes: green dots and PBC (black dots) eigenspectra in the complex energy plane $Re(E)-Im(E)$ where $E=E_{\pm}\in \mathbb{C}$ as in Eq. \ref{eq:model_eq_eig} for the black dots and $E=E_{OBC}\in \mathbb{C}$ for blue, red, and green dots.}
    \label{fig8} 
\end{figure*}

In fact, it is quite clear from these plots in the case of weak lossy non-Hermiticity that Fig. \ref{fig6}(a1-a2) show predominantly extended modes ($n_E \gg n_{L,R}$) with few-number of left- and right-localized \textit{edge modes}, followed by decreasing number of extended modes in Fig. \ref{fig6}(b1-b2) but with mixed number of left- and right-localized skin modes. Hence, it is indicated as “mostly extended” as shown in Fig. \ref{fig6}(a1). The number of extended modes keeps decreasing further and the numbers of left- and right-localized skin modes start to equalize as $n_L\approx n_R, n_E \neq 0$ from Fig. \ref{fig6}(c1-c2). We have indicated this as ‘mixed’ in Fig. \ref{fig6}(b1),(c1). In Fig. \ref{fig6}(d1,f1), the number of extended modes has decreased further and the numbers of left- and right-localized skin modes have almost become equal but $n_E \neq 0$. Hence, we have indicated these two cases as “dipole-like.” Finally, in Fig. \ref{fig6}(e1-e2), the number of extended modes is zero and all modes are equally distributed as left- and right-localized skin modes with $n_L\approx n_R, n_E = 0$. Therefore, in Fig. \ref{fig6}(e1), we have indicated this as reminiscent of the “pure dipole” skin effect. For strong lossy non-Hermiticity, excitation of a few number of extended modes occurs $n_L\approx n_R, (n_L, n_R)\gg n_E$.

In terms of the position of the extended mode eigenvalues, they reside on the left- or right-side open-arc-like OBC spectra, and at a given value of $\theta_{L}^{NNN}=-2.15$ (when other parameters are given as above), all extended mode eigenvalues disappear leaving the OBC spectra represented only by the eigenvalues corresponding to the left- and right-localized eigenstates. We may note that this open-arc-like conjoint OBC spectrum occurs when NNN hopping amplitudes are held fixed at their reciprocal limits, but the non-Abelian phase is varied over a range $\theta_{L}^{NNN}\in (-1.0,2.2)$. Beyond these values, the conjoint open-arc structures gradually deform. They do not appear when $\theta_{L}^{NNN}$ and $\theta_{R}^{NNN}$ are fixed at their reciprocal limits, thus signifying their strong non-Hermitian origin because of the non-Abelian phases. If we keep all other parameters fixed at $\gamma=0.0,$ $J_{L}^{NN}=J_{R}^{NN}=1.0,$ $\theta_{L}^{NN}=\theta_{R}^{NN}=0.0,$ $\theta_{L}^{NNN}=-2.1,$ $\theta_{R}^{NNN}=-1.0,$ and vary the values of $J_{L}^{NNN}$ and $J_{R}^{NNN}$ while keeping them at their reciprocal limits, a parametric window is found as $\Delta J_{L,R}^{NNN}\in(0.48,0.85)$ where the conjoint OBC spectra are intact. The conjoint open-arc OBC structure begins to form when $J_{L}^{NNN}=J_{R}^{NNN}\approx 0.26$ for the above set of parameter values, as an example. Below this value, the OBC spectra remain conjoint but with a deformed open-arc structure. If we denote by $J_{L,R}^{NNN,max}$ the maximum value of $\alpha_{1}$ $(J_{L}^{NNN}=J_{R}^{NNN}=\alpha_{1}$) and $J_{L,R}^{NNN,min}$ the minimum value of $\alpha_{2}$ $(J_{L}^{NNN}=J_{R}^{NNN}=\alpha_{2}),$ then $(J_{L,R}^{NNN,min},J_{L,R}^{NNN,max})=(0.48, 0.85).$ It may be noted that the conjoint open-arc OBC structure is still found when $0.85\leq J_{L,R}^{NNN,max}\leq 1.0.$ We consider $J_{L,R}^{NNN,max}=0.85$ in order to keep it below the NN hopping amplitudes. In summary, in the formation of the conjoint open-arc structure, a number of distinct stages are noticed. \textit{Stage 1}: for the above set of parameters, when $J_{L}^{NNN}=J_{R}^{NNN}\leq0.22$ coexisting left- and right-localized skin modes, and extended delocalized modes occur with a hybrid skin effect. Beyond this value, the open-arc structure begins to appear that becomes prominent near $J_{L}^{NNN}=J_{R}^{NNN}\approx0.26.$ At this stage, the OBC spectrum is still hybrid with a number of extended modes gradually decreasing. \textit{Stage 2}: for the same set of parameters, increasing $J_{L}^{NNN}=J_{R}^{NNN}$ beyond 0.26 results in the gradual formation of the complete conjoint OBC spectra with the decreasing number of extended modes. When $J_{L}^{NNN}=J_{R}^{NNN}\approx0.40$ the conjoint open-arc OBC spectrum is complete with all eigenmodes being skin modes localized to the left and right. \textit{Stage 3}: for the same set of parameter values for $J_{L}^{NNN}=J_{R}^{NNN}\geq0.40$ the conjoint open-arc OBC structure remains robust unless other parameters change. 
In addition, we note another interesting point about the conjoint open-arc OBC spectra. It is fairly robust with respect to the change in $\gamma$, while other parameters remain the same (see Fig. \ref{fig10_appen}(d)). As long as $\gamma$ remains weak or moderate, the shape of the conjoint open-arc OBC spectra does not change and remain entirely populated by the left- and right-localized skin mode eignevalues. For high values of $\gamma$, only one or two extended modes are excited. In addition, the localization dynamics of the OBC eigenstates and existence of conjoint open-arcs remain unchanged with increasing system size $N$ (see Appendix \ref{AppendixB}). On the other hand, it is quite sensitive to other parameters, such as non-Abelian NNN phases $\theta_{L}^{NNN}, \;\theta_{R}^{NNN}$. Changing it slightly induces the emergence of the delocalized extended modes, disrupting the conjoint open-arc structure. Therefore, gain/loss non-Hermiticity plays a less prominent role than non-Hermiticity resulting from the non-Abelian phases. A potential question that may arise is whether there is any underlying connection between the \textit{pure} dipole skin effect and the conjoint open-arc OBC spectra? Our analysis suggests that there, in fact, could be an underlying correspondence between the \textit{pure} dipole skin effect and the occurrence of the conjoint open-arcs in our model. Moreover, it could be worthwhile to note the apparent distinction between the single open-arc structure \cite{pang2024synthetic} and the conjoint open-arc spectra reported in this work. 

We now turn our attention to the case of alternately distributed gain-loss non-Hermiticity across the whole lattice as depicted in Fig. for the same set of parameters as in Fig. \ref{fig6}. We find that Fig. \ref{fig7}(a1-a2) show predominantly extended modes ($n_E \gg n_{L,R}$) with few-number of left- and right-localized defect \textit{edge modes}, followed by decreasing number of extended modes in Fig. \ref{fig7}(b1-b2) but with mixed number of left- and right-localized skin modes. Hence, we have indicated this by “Mostly extended” in Fig. \ref{fig7}(a1-a2). The number of extended modes keeps decreasing further and the numbers of left- and right-localized skin modes start to equalize as $n_L\approx n_R, n_E \neq 0$ in Fig. \ref{fig7}(c1-c2), (d1-d2). We have indicated this as ‘mixed’ both in Figs. \ref{fig7}(c1-c2) and Fig. \ref{fig7}(d1-d2). In Figs. \ref{fig7}(e1-e2) and Figs. \ref{fig7}(f1-f2), the number of extended modes has further decreased and the numbers of left- and right-localized skin modes have almost become equal but $n_E \neq 0$. Hence, we have indicated these two cases as “dipole-like.” In sharp contrast to Fig. \ref{fig7}(e1-e2), we note that the number of extended modes is never zero although the left- and right-localized skin modes appear to form the "dipole-like" skin effect. Therefore, in this case, the “\textit{pure} dipole” skin effect is absent.

For the sake of completeness of our analysis, we have considered the case when there is no gain-loss distributions in the lattice, $\textit{i.e.}$ $\gamma =0$. In fact, it is quite clear from these plots that Fig. \ref{fig8}(a1-a2) show completely extended modes (here, $n=n_E , n_{L}=n_{R}=0$ in sharp contrast to the "mostly extended" case with excitation of few defect edge modes as in Fig. \ref{fig6}(a1-a2) with $n_{E}\neq0$), followed by decreasing number of extended and increasing numbers of left- and right-localized skin modes in Fig. \ref{fig8}(b1-b2), (c1-c2),(d1-d2) but with mixed number of left- and right-localized skin modes. Hence, we have indicated this by “completely extended” in Fig. \ref{fig8}(a1) and ‘Mixed’ both in Fig. (b1-b2), (c1-c2),(d1-d2). In Figs. \ref{fig8}(e1-e2) and Figs. \ref{fig8}(f1-f2), the number of extended modes has further decreased and the numbers of left- and right-localized skin modes have almost become equal. Out of these, we have indicated the distribution in Fig. \ref{fig8}(e1-e2) as “\textit{pure } dipole" skin effect since here $n_L \approx n_R$ and $n_E=0$. Finally, in Fig. \ref{fig8}(f1-f2), the number of extended modes is highly diminished and all modes are equally distributed as left- and right-localized skin modes with $n_L\approx n_R, n_E = 0$. Therefore, in Fig. \ref{fig8}(f1-f2), we have indicated this as reminiscent of the “dipole-like” distribution. 

\section{\label{sec:level1} Conclusion and discussion}
A new class of synthetic photonic lattices is studied with simultaneous considerations of NN and NNN hopping interactions, non-Abelian SU (2) gauge phases, and gain/loss non-Hermiticity. These multiple competing sources of non-Hermiticity induce novel physical effects, such as EP-mediated non-Hermitian topological phase transition of an intertwining nested Hopf-link structure and the conjoint open-arc OBC spectra with a \textit{pure} dipole skin localization. It is shown how increasing gain/loss non-Hermiticity beyond EP leads to trivialization of the non-Hermitian bands post an imaginary gap opening. The topological phase of the nested Hopf-link structure is quantified by a topological invariant known as the linking number of the Gauss' integral form. The non-Hermitian topological phase transition of the nested Hopf-link structure is demonstrated through the calculation of the topological linking number that undergoes a nontrivial to trivial topological transition at the EP. A conjoint open-arc OBC structure is observed that corresponds to the solely left and right skin localizations without any extended modes, thus illustrating the plausible underlying connection between the conjoint open-arc OBC spectra and a \textit{pure} dipole skin localization. 

\begin{figure*}
    \centering
    \includegraphics[width=0.99\linewidth]{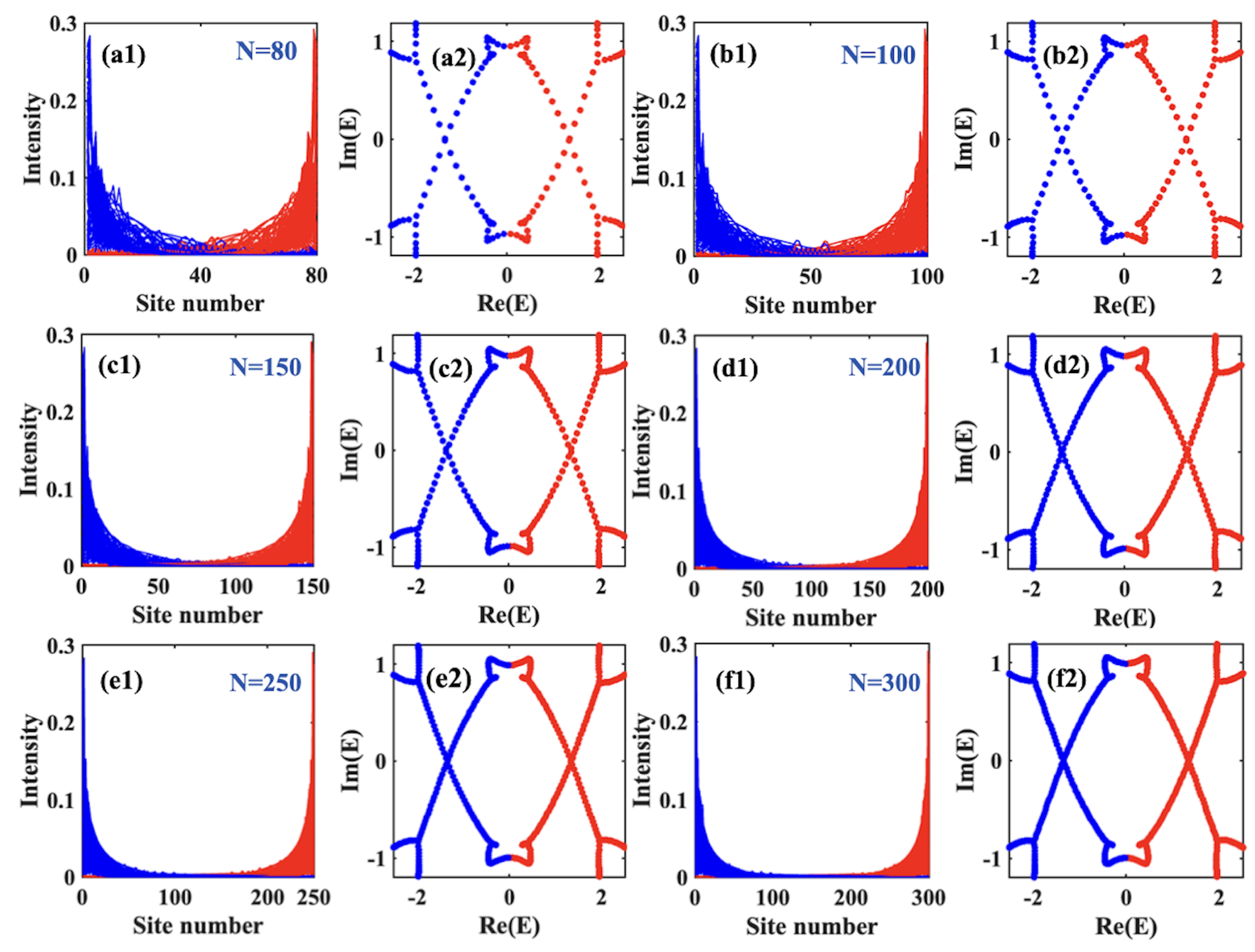} 
    \caption{Effects of system size $N$ on the formation of the \textit{pure} dipole-skin effect and corresponding conjoint open-arcs as shown in Fig. \ref{fig6}(e1-e2) for $\theta_{L}^{NNN}=-2.15$ showing its robust persistent nature. Other parameters: $J_{L}^{NN}=J_{R}^{NN}=1.0$, $J_{L}^{NNN}=J_{R}^{NNN}=0.85$, $\theta_{L}^{NN}=\theta_{R}^{NN}=0.0$,\, $\theta_{R}^{NNN}=-1.0$ and $\gamma=0.2$. In all these panels, the subfigures (a1,b1,c1,d1,e1,f1) depict the intensity distributions of the OBC eigenstates across the lattice sites $N$ (left-localized modes ($n_L$): blue curves, right-localized modes ($n_R$): red curves, extended modes ($n_E=0$): green curves) and the subfigures (a2,b2,c2,d2,e2,f2) show the corresponding  OBC (left-localized modes: blue dots, right-localized modes: red dots, extended modes: green dots (($n_E=0$)) in the complex energy plane $Re(E)-Im(E)$ where $E=E_{OBC}\in \mathbb{C}$. We see that with increasing value of $N$ sharper left- and right skin localization takes place with the OBC complex spectrum remaining unchanged throughout, as expected.}
    \label{fig9_appen} 
\end{figure*}

\begin{figure*}
    \centering
    \includegraphics[width=0.99\linewidth]{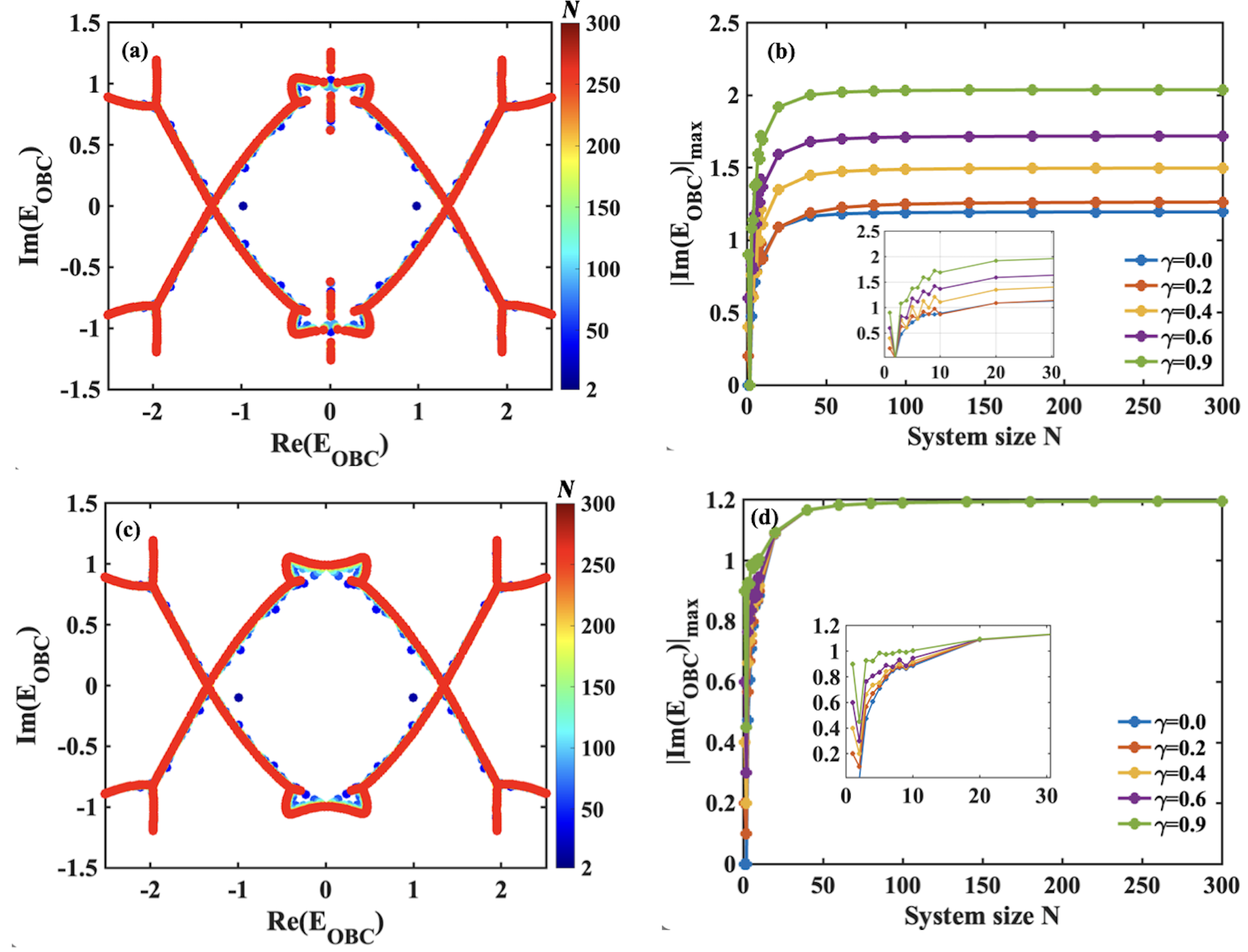} 
    \caption{Evolution of the OBC eigenspectra $E=E_{OBC}\in \mathbb{C}$ in the complex plane with $\gamma=0.2$ and different system size $N$ (a,b) and maximum of the OBC eigenenergy for different values of $\gamma$ (c,d) of the conjoint open-arcs shown in Fig. \ref{fig7}(e1-e2) and Fig. \ref{fig6}(e1-e2) corresponding to the alternately distributed gain-loss and lossy end-site non-Hermiticity cases respectively. Here, (a,b) show spectral convergence of the OBC eigenspectra and (c,d) show the robust persistent nature of the \textit{pure} dipole (for lossy end-non-Hermiticity case) and dipole-like skin effects and conjoint open-arcs. The insets in (c,d) show the zoomed in views of the maximum imaginary parts of the eigenvalues. Other parameters: $J_{L}^{NN}=J_{R}^{NN}=1.0$, $J_{L}^{NNN}=J_{R}^{NNN}=0.85$, $\theta_{L}^{NN}=\theta_{R}^{NN}=0.0$,\, $\theta_{R}^{NNN}=-1.0$, $\theta_{L}^{NNN}=-2.15$, $\theta_{R}^{NNN}=-1.0$.}
    \label{fig10_appen} 
\end{figure*}

\begin{figure*}
    \centering
    \includegraphics[width=0.99\linewidth]{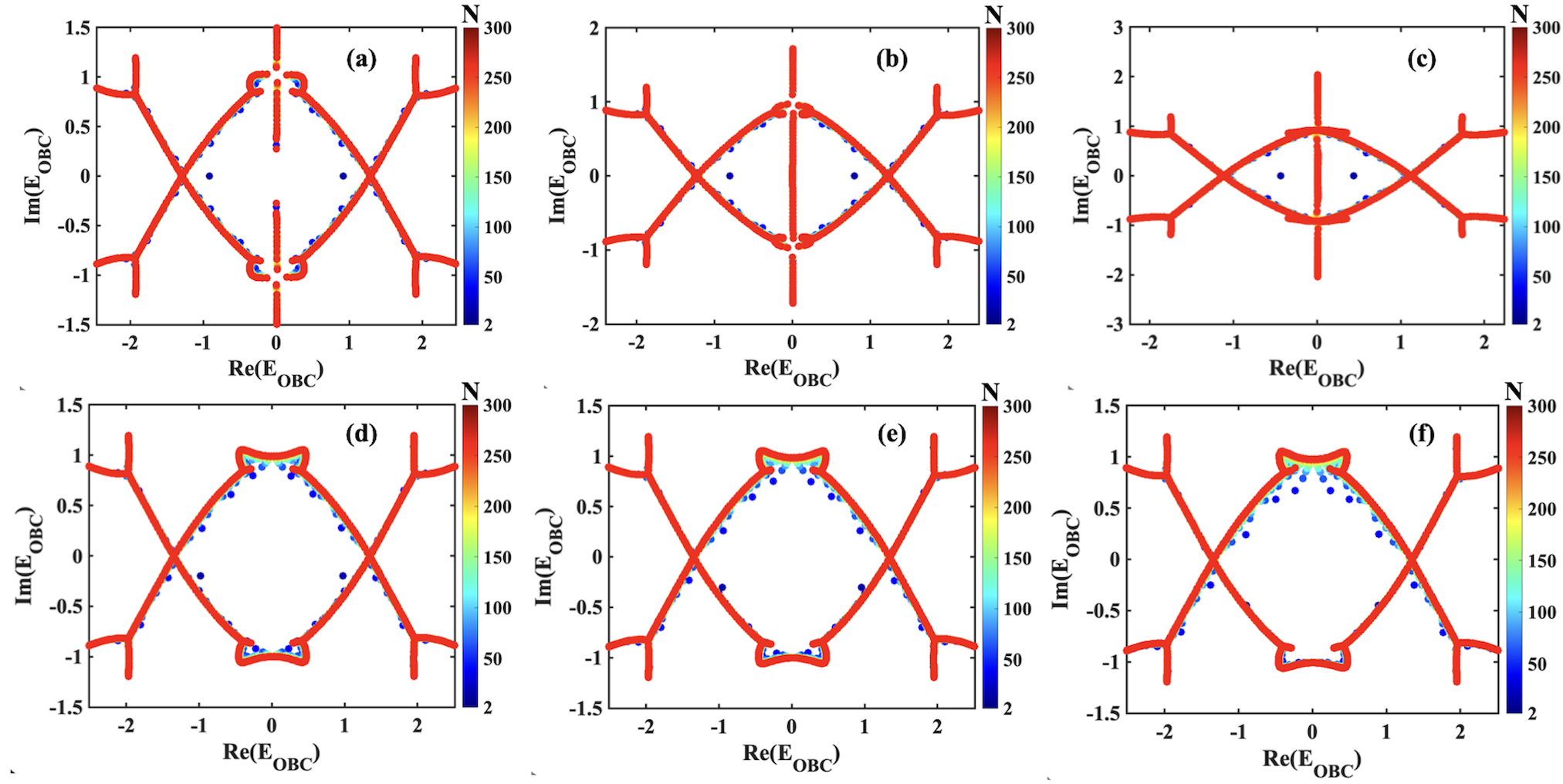} 
    \caption{Evolution of the OBC eigenspectra $E=E_{OBC}\in \mathbb{C}$ in the complex plane with varying $\gamma$ (in addition to $\gamma=0.2$ in Fig. \ref{fig10_appen}(a,c)) and different system size $N$ of the conjoint open-arcs shown in Fig. \ref{fig7}(e1-e2) and Fig. \ref{fig6}(e1-e2) corresponding to the alternately distributed gain-loss and lossy end-site non-Hermiticity cases respectively. The gain-loss distributed lattice: (a) $\gamma=0.4$, (b) $\gamma=0.6$, (c) $\gamma=0.9$. The end-site non-Hermiticity case: (d) $\gamma=0.4$, (e) $\gamma=0.6$, (f) $\gamma=0.9$. Here, for larger values of $N$ (a,b,c) show that the OBC spectra undergo structural change and (d,e,f) show the robust persistent nature of the OBC spectral structures corresponding to the \textit{pure} dipole effect and conjoint open-arcs. Other parameters: $J_{L}^{NN}=J_{R}^{NN}=1.0$, $J_{L}^{NNN}=J_{R}^{NNN}=0.85$, $\theta_{L}^{NN}=\theta_{R}^{NN}=0.0$,\, $\theta_{R}^{NNN}=-1.0$, $\theta_{L}^{NNN}=-2.15$, $\theta_{R}^{NNN}=-1.0$.}
    \label{fig11_appen} 
\end{figure*}

\begin{figure*}
    \centering
    \includegraphics[width=0.99\linewidth]{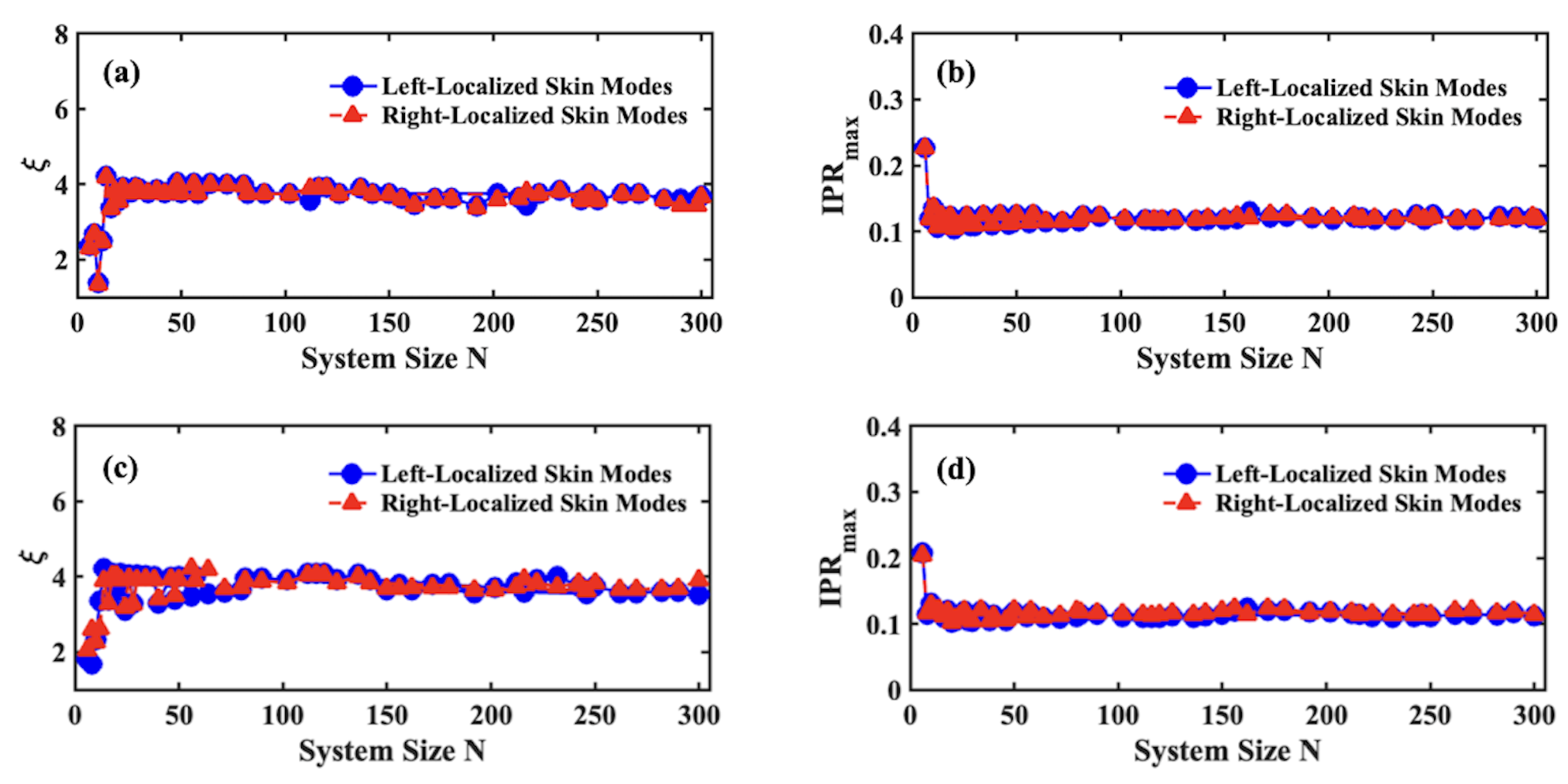} 
    \caption{Localization length (a,c) and inverse participation ratio (IPR) (b,d) for the left- and right-localized skin modes for different values of system size $N$ corresponding to the alternately distributed gain-loss and lossy end-site non-Hermiticity cases as shown in Fig. \ref{fig7}(e1-e2) and Fig. \ref{fig6}(e1-e2). Here, in (a,c) the localization lengths are $\xi_{left}\approx\xi_{right}=3.70$ and $\xi_{left}\approx3.53, \xi_{right}\approx3.91$, and in (b,d) the maximum of IPR are $IPR^{max}_{left}\approx IPR^{max}_{right}=0.119$ and $IPR^{max}_{left}\approx 0.112, IPR^{max}_{right}\approx0.114$. For small $N$, effect of finite-size can be seen as expected in the initial transient growths. The plots strongly indicate size-independent \textit{pure} dipole and dipole-like skin localization for these cases. show the robust persistent nature of the \textit{pure} dipole (for lossy end-site non-Hermiticity case) and dipole-like skin effects (alternately gain-loss lattice) and conjoint open-arcs. The insets in (c,d) show the zoomed in views of the maximum imaginary parts of the eigenvalues. Other parameters: $J_{L}^{NN}=J_{R}^{NN}=1.0$, $J_{L}^{NNN}=J_{R}^{NNN}=0.85$, $\theta_{L}^{NN}=\theta_{R}^{NN}=0.0$,\, $\theta_{R}^{NNN}=-1.0$, $\theta_{L}^{NNN}=-2.15$, $\theta_{R}^{NNN}=-1.0$.}
    \label{fig12_appen} 
\end{figure*}

It is worth mentioning that the \textit{pure} dipole skin effect in our work is distinct in origin from conventional topologically protected non-Hermitian skin effects driven solely by non-reciprocity \cite{okuma2020topological, zhang2022review, zhu2020photonic,okuma2023non, lin2023topological}. In our model, the pure dipole skin effect emerges from the interplay of non-Abelian SU(2) gauge phases and NNN hopping, and is directly tied to the formation of a conjoint open-arc structure in the OBC spectrum. Moreover, it is robust over a small to medium range of onsite disorder and is likely to be experimentally realized (see Appendix \ref{AppendixC}).

The coexisting left- and right-localized skin modes discovered in this system share physical similarities with the well-known bipolar non-Hermitian skin effect \cite{yuce2024strong, jiang2023reciprocating, rafi2024twisted, jiang2024tunable, zou2025bipolar}. However, the "pure dipole skin effect" highlighted in this work represents a distinct, extreme regime of the bipolar skin effect. Driven by the specific geometric formation of the conjoint open-arc OBC spectra under NNN non-Abelian gauge phases, our system filters out the delocalized bulk continuum, completely funneling the eigenstates into localized edge distributions without bulk leakage. This stands in contrast to typical bipolar skin effects where a portion of extended or mixed bulk modes typically persists.

We emphasize that there is an apparent distinction between the nested Hopf-link structures reported in this work and traditional Hopf phases. The former refer to linked trajectories of complex energy eigenvalues in the complex energy plane as the quasimomentum $k$ is varied over the Brillouin zone. In contrast, the latter arise mostly in three-dimensional systems, where a topological Hopf invariant is defined through a continuous mapping from the three-dimensional Brillouin zone to the Bloch sphere \cite{liu2017symmetry, lapierre2021n}.

While the work in \cite{midya2023gain} demonstrates gain-loss-induced braid transitions in multiband lattices with Hermitian hoppings, and the work in \cite{pang2024synthetic} establishes the existence of Hopf-link bulk braiding and coexisting skin modes in a basic nearest-neighbor (NN) non-Abelian Hatano-Nelson model, the present work introduces a highly generalized configuration featuring multiple competing non-Hermitian degrees of freedom. Specifically, by simultaneously incorporating next-nearest-neighbor (NNN) hopping interactions, non-Abelian SU(2) gauge phases, and a staggered onsite gain-loss profile, our system yields a number of unique physical features. These include the formation of intertwining nested Hopf-link structures in the bulk spectrum undergoing a EP-mediated topological phase transition and the emergence of a conjoint open-arc open boundary condition (OBC) spectrum with corresponding bipolar skin localization. This spectral topology results in a pure dipole skin effect where the entire continuum of bulk states completely transforms into edge-localized modes without any surviving delocalized extended states.

Exploring non-Hermitian topology of such synthetic non-Abelian systems is relatively a new idiosyncratic avenue of research. In the future, a number of potential questions could be worth exploring. For example, the model considered here is a simplified version of a more generic class of complex photonic settings. These generic settings may require numerical methods to track the relevant dynamics of exceptional points, photonic exceptional structures, and various topological phases. It would be interesting if further studies are carried out in more generic models where a set of new physical insights owing to the complex interaction of non-Hermitian skin localization and non-Abelian dynamics are expected to emerge in the presence of other natural and engineered factors (such as disorder, long-range hoppings, or complexified extension of SU(2) gauge fields) in such systems. 

The demonstrations of conjoint open arcs, pure dipole skin effects, and EP-mediated topological phase transitions may lead to novel applications in wave-steering platforms. For example, by judicious tailoring of the OBC eigenspectrum into conjoint arcs, the system completely filters out extended bulk states, funneling the entire continuum of eigenstates into a localized dipole skin distribution without leakage into the bulk. This robust, complete localization can be expected to be directly engineered in synthetic gauge platforms such as synthetic frequency lattices \cite{wang2021topological}, time-multiplexed photonic lattices \cite{parto2023non}, topolectrical circuits \cite{chen2026implementing}, and acoustic metamaterials \cite{zhang2025experimental} to construct high-efficiency multi-channel amplifiers, directional routers and sensors, robust topological switches, non-reciprocal laser amplifiers, and fault-tolerant waveguiding devices. Therefore, this may stimulate further research activities at the intersection between non-Hermitian and non-Abelian topological phenomena, and find applications in robust and fault-tolerant communication channels, computing platforms, and non-Hermitian topological devices \cite{wang2021topological, zhang2022non, yang2024non,yang2019synthesis,sun2025reconfigurable}.
\begin{acknowledgments}
S.K.G. thanks the Department of Physical Sciences, IISER Berhampur for support through the Postdoctoral Research Fellowship.
\end{acknowledgments}
\section*{DATA AVAILABILITY}
The data are not publicly available. The data are available from the authors upon reasonable request.

\appendix
\section{Topological linking number}\label{AppendixA}
The topological linking between two closed curves $C_{+}$ and $C_{-}$ can be quantified by the linking number of the Gauss' integral form, which is a topological invariant measuring the number of times one curve winds around the other. It is defined as \cite{yang2019non, kim2023realization, ricca2011gauss}:
\begin{equation}
\mathrm{L}(C_{+},C_{-})=
\frac{1}{4\pi}
\oint_{C_{+}}
\oint_{C_{-}}
\frac{
\left(\mathbf{r}_{+}-\mathbf{r}_{-}\right)
\cdot
\left(
d\mathbf{r}_{+}
\times
d\mathbf{r}_{-}
\right)
}
{\left|\mathbf{r}_{+}-\mathbf{r}_{-}\right|^{3}},
\label{eq:GaussLink}
\end{equation}
with $\mathrm{L}(C_+,C_-)=\mathrm{L}(C_+,C_-)$ where $\mathbf{r}_{+}$ and $\mathbf{r}_{-}$ denote the position vectors of the points on the closed curves $C_{+}$ and $C_{-}$, respectively. The infinitesimal line elements $d\mathbf{r}_{+}$ and $d\mathbf{r}_{-}$ are tangent vectors along the respective curves. The vector difference $\mathbf{r}_{+}-\mathbf{r}_{-}$ joins two infinitesimal segments on the curves, while in the denominator $|\mathbf{r}_{+}-\mathbf{r}_{-}|$ represents the straight-line distance between the two points on the curves. 
In our case, the complex energy eigenvalues are given as:
\begin{equation}
E_{\pm}(k)= Re(E_{\pm}(k))+i\,Im(E_{\pm}(k)).
\end{equation}
The spectral trajectories of the curves can be embedded into a 3D parameter space spanned by:
\begin{equation}
\mathbf{r}_{\pm}(k)=(Re(E_{\pm}(k))\,,Im(E_{\pm}(k)),\,k).
\end{equation}

The Gauss linking number is an integer-valued topological invariant for two non-intersecting closed curves and remains unchanged under continuous deformations that preserve their topology. In non-Hermitian band theory, the curves $C_{+}$ and $C_{-}$ may represent exceptional lines, nodal loops, or trajectories of complex-energy bands in momentum space. Consequently, $\mathrm{L}=0$ corresponds to unlinked curves, whereas $\mathrm{L}=\pm1,\pm2,\ldots$ characterize increasingly intricate linked topological configurations. This construction is particularly well suited for non-Hermitian systems because it treats each complex-energy band as a closed space curve in the augmented parameter space $(Re(E_{\pm}(k))\,,Im(E_{\pm}(k)),\,k)$. A nonzero linking number indicates that two spectral trajectories cannot be continuously disentangled without closing the point-gap or undergoing a topological phase transition, thereby providing a robust topological characterization of linked exceptional structures and braided complex-energy bands.

\section{Effects of system size}\label{AppendixB}
To examine the effects of system size on OBC eigenspectra and localization profiles of OBC eigenstates, we repeated numerical calculations for different system sizes $N$. We find that the qualitative structure of the OBC eigenspectra, and the localization profiles of the OBC eigenstates remain essentially unchanged as the system size increases. For example, the overall structures of the OBC spectra remain unchanged, conjoint open-arc spectral structure remains the same, and localization profiles of the left- and right-localized skin modes remain robust. 

One example is directly shown in Fig. \ref{fig9_appen} where the existence of the conjoint open-arcs and related localization profiles of the left- and right-localized skin modes, \textit{i.e.} \textit{pure} dipole skin effect as shown in Fig. \ref{fig6}, reveal the size-independent responses with respect to the system size $N$. Similar conclusions can be drawn for the \textit{pure} dipole skin effect and conjoint open-arcs shown in Fig. \ref{fig8}(e1-e2), and nearly dipole-like skin effects and conjoint open-arcs in Fig. \ref{fig7}(e1-e2), (f1-f2) (not shown for the sake of brevity).

In order to investigate the robustness of the conjoint OBC spectra for the alternately gain-loss distributed and end-site non-Hermiticity cases with respect to system size $N$, the OBC energy eigenpectra and maximum of the imaginary parts of the eigenvalues are illustrated in Fig.\ref{fig10_appen}(a,c) and (b,d) respectively. It shows that the OBC eigenspectra remain unchanged over a wide range of values of $N$ in both cases. In addition, after initial finite-size transient response, the maximum of the imaginary parts of the OBC eigenvalues show saturation behaviors beyond some value of $N$. All these curves corresponding to different values of $\gamma$ remain well-separated. Upon a close inspection, it is revealed that in Fig.\ref{fig10_appen}(b) for $N=2$ there is a complex to real, and real to complex phase transitions (for $N=2$ there is a pair of isolated purely real eigenvalues lying on $Im(E_{OBC})=0)$. However, this never occurs for the case of Fig.\ref{fig10_appen}(d) and the curves are found to coincide and remain coincident at $N\approx30$ and onward. This affects the structures of the OBC eigenspectra with varying numbers of $N$ as $\gamma$ is varied. Here, we find that the structure of the OBC spectrum remains conjoint although it undergoes qualitative changes as $N$ increases for the alternately gain-loss lattice (Fig. \ref{fig11_appen}(a,b,c)) while they remain unchanged for the end-site non-Hermiticity case (Fig. \ref{fig11_appen}(d,e,f)) as the onsite gain-loss disorder parameter $\gamma$ is varied to higher values.

Furthermore, the localization length ($\xi$) and the inverse participation ratio (IPR) have been depicted in Fig. \ref{fig12_appen} for the alternately gain-loss lattice and end-site non-Hermiticity cases to show system size-independent behaviors. Both localization length and IPR show system size-independent behaviors for higher values of $N$ where they both saturate to constant values. The IPR turns out to be 0.119 for the alternate gain-loss lattice, and 0.112 (left) and 0.114 (right) for the end-site non-Hermiticty cases, respectively. This is in accordance with the dipole-like and pure dipole skin localizations as previously shown in Fig. \ref{fig7}(e1) and Fig. \ref{fig6}(e1), respectively.

\section{Disorder robustness}\label{AppendixB}

In the tight-binding second-quantized representation, the disorder Hamiltonian accounting for the sublattice basis $\alpha \in \{A, B\}$ is given by:
\begin{figure*}
    \centering
    \includegraphics[width=0.9999\linewidth]{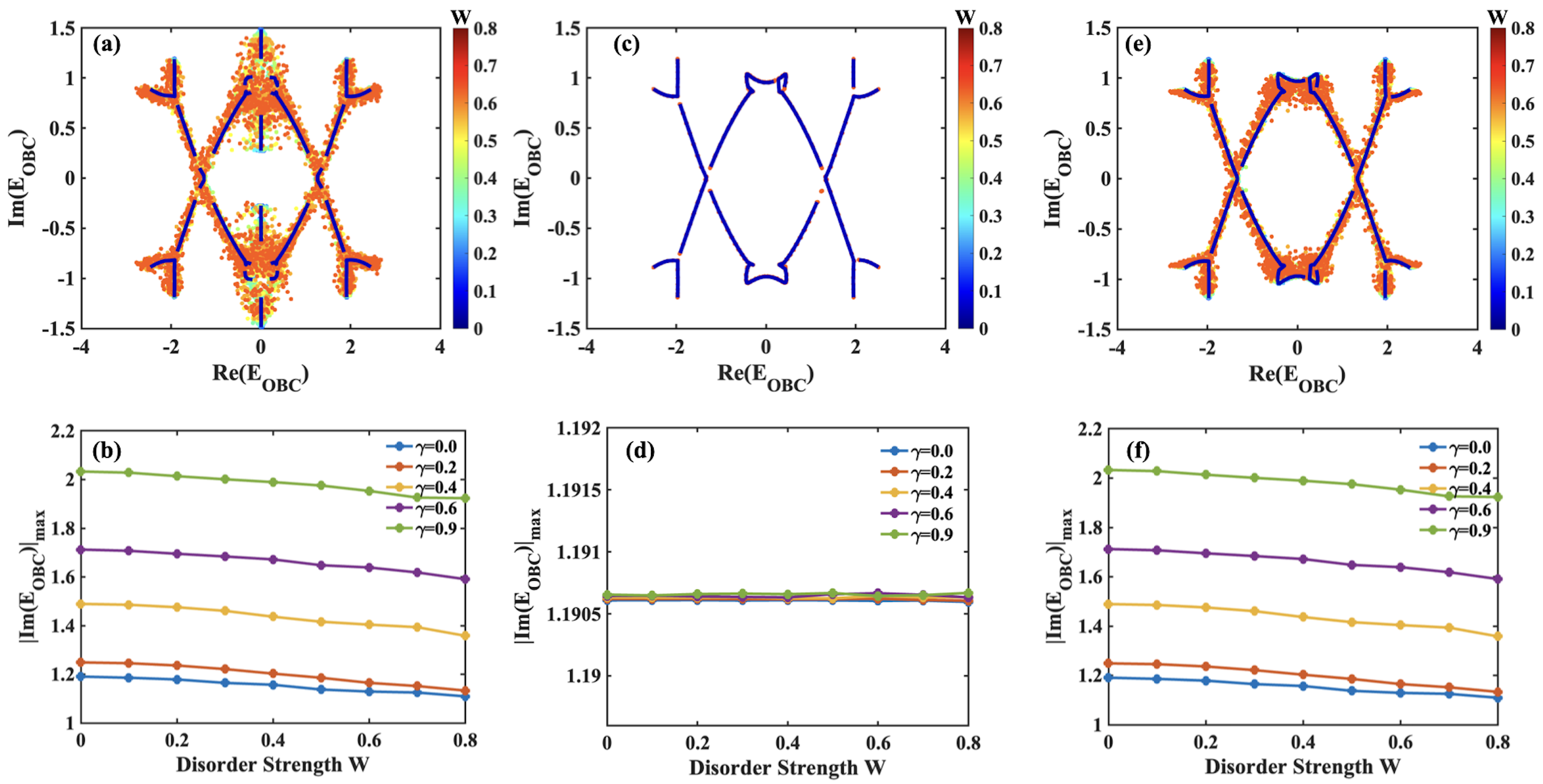} 
    \caption{Disorder robustness of the conjoint open arcs with respect to real onsite disorder $\epsilon_{n}$. (a,b) For the gain-loss distributed ($\gamma=0.4$) lattice where (a) shows the OBC eigenspectra $E=E_{OBC}\in \mathbb{C}$ in the complex energy plane $Re(E)-Im(E)$ varying onsite real disorder $W$ and (b) shows the $|Im(E_{OBC})|_{max}$ vs disorder strength $W$. The same plots are shown in (c,d) for the cases of end-site non-Hermiticity ($\gamma=0.4$) and in (e,f) for the no gain/loss case with $\gamma=0.0$. Other parameters: $J_{L}^{NN}=J_{R}^{NN}=1.0$, $J_{L}^{NNN}=J_{R}^{NNN}=0.85$, $\theta_{L}^{NN}=\theta_{R}^{NN}=0.0$,\, $\theta_{R}^{NNN}=-1.0$, $\theta_{L}^{NNN}=-2.15$, $\theta_{R}^{NNN}=-1.0$.}
    \label{fig12_dis} 
\end{figure*}
\begin{equation}
    H_{\text{dis}} = \sum_{j=1}^{N} \sum_{\alpha \in \{A, B\}} \epsilon_{j, \alpha} \, c_{j, \alpha}^\dagger c_{j, \alpha}
\end{equation}
where $c_{j, \alpha}^\dagger$ ($c_{j, \alpha}$) denotes the creation (annihilation) operator for an unit cell $j$ on sublattice $\alpha$. The onsite fluctuations $\epsilon_{j, \alpha}$ are independent, uniformly distributed random variables sampled from the interval $[-W, +W]$, where $W$ parameterizes the overall disorder strength.
Because the disorder is purely localized on individual sites and does not induce spin flips on its own, it acts as a scalar identity operator in the internal pseudo-spin space. This requires the following disorder term to be included in the Hamiltonian in the OBC case $\epsilon_{dis}=\epsilon_n * \sigma_0$, where $\epsilon_n=W\times (2\times rand(1)-1)$ and the function rand(1) generates random numbers in the range of [0,1] so that $\epsilon_n \in [-W,W]$. Investigation of disorder in non-Hermitian lattices is common \cite{hatano1998non, komis2026evolving, longhi2025lifshitz}. In realistic setups like photonic chips or coupled ring resonators, an imaginary staggered potential corresponds to engineered amplification and attenuation rates. However, any physical fabrication process inevitably introduces random structural imperfections, such as, slight variations in ring radii or waveguide widths. These imperfections manifest as random shifts in the local resonant frequencies, which act as real onsite disorder.

Here, in Fig. \ref{fig12_dis} we have shown the disorder robustness of conjoint open arcs under the influence of real onsite potential disorder. In doing so, we calculate both the complex eigenvalue distribution and the ensemble-averaged maximum imaginary energy scaling, $|Im(E_{OBC})|_{max}$.

In the zero disorder limit ($W=0$) as shown in Fig. \ref{fig12_dis} (a,c,e), the OBC spectrum maps onto a highly symmetric, conjoint open arc structure represented by the solid blue line. In Fig. \ref{fig12_dis} (c), the conjoint open arc remains highly robust with respect to disorder with flat $|Im(E_{OBC})|_{max}$ vs W curves shown in (d). In Fig. \ref{fig12_dis} (a,e), the introduction of a real onsite disorder causes the individual eigenvalues to undergo local fluctuations, generating slightly broadened spectral distributions that expand outward with increasing disorder strength up to ($W=0.8$). Crucially, this smearing does not trigger a structural collapse of the spectrum. The eigenvalues remain strictly confined and tightly pinned along the disorder-free reference trajectories of the conjoint open arcs without merging or closing.

This disorder robustness of the conjoint open arc spectra is further corroborated by the scaling behavior of $|Im(E_{OBC})|_{max}$ across varying gain-loss parameters $\gamma$. For (b) and (f), the maximum imaginary energy component exhibits a remarkably gradual, nearly-linear decrease as the disorder strength increases while for (d), these curves are almost coincident. The absence of steep decrease and abrupt discontinuities demonstrates that the system is protected from abrupt delocalization or other phase transitions. While higher baseline gain-loss parameters expectedly scale the overall imaginary energy bounds upward, the decay rates remain remarkably uniform and stable across all curves.

On a physical level, this robust behavior of the conjoint open arcs highlights the interplay between non-Hermitian skin dynamics and non-Abelian SU(2) gauge fields. The underlying dipole skin localization remains topologically protected, forcing the states to resist local potential fluctuations. These findings confirm that the conjoint open arcs are not fragile artifacts of such lattice models, but robust topological features capable of surviving realistic fabrication imperfections, potentially rendering this platform as a viable candidate for stable, fault-tolerant non-Hermitian photonic and quantum optical devices.

\section{\label{AppendixC}Experimental feasibility}
The judicious preparation of the experimental settings and successful demonstration of the observed physical effects are based on the effective control of three primary physical processes as described in the following.
\subsection{Non-Abelian synthetic gauge fields}
The implementation of non-Abelian gauge fields in photonic systems has witnessed remarkable progress in recent years, establishing synthetic gauge fields as a practical tool for engineering topological light transport. Initial experimental demonstrations realized synthetic SU(2) non-Abelian gauge fields for classical optical waves using anisotropic metamaterials, where polarization degrees of freedom emulate pseudospin states and the associated non-Abelian gauge potentials give rise to optical analogs of the Aharonov--Bohm effect and Wilson-loop observables \cite{chen2019non, yang2019synthesis}. More recently, dynamically modulated photonic synthetic-frequency lattices based on coupled ring resonators have enabled the experimental realization of non-Abelian lattice gauge fields, in which the photon polarization acts as an internal spin degree of freedom while modulation-induced inter-frequency couplings implement matrix-valued SU(2) hopping phases \cite{cheng2025non}. These platforms provide direct access to non-Abelian gauge physics and topological band engineering in programmable photonic lattices. In parallel, integrated photonic chips have experimentally demonstrated non-Abelian braiding and holonomic operations through tunable waveguide networks, further confirming the feasibility of engineering controllable non-commuting gauge transformations in optical systems \cite{zhang2022non,sun2025reconfigurable, jiang2026photonic}. In particular, a two-mode SU(2) internal degree of freedom is achievable through polarization modes, dual spatial modes, or two-resonator “photonic molecules” within each unit cell, as experimentally demonstrated in non-Abelian photonic braiding on-chip by judiciously controlling the geometric-phase matrix in photonic waveguide arrays \cite{zhang2022non}. More recently, it has been proposed that synthetic photonic SU(2) pseudospin can be achieved in free space optics and integrated photonicc systems \cite{wong2025synthetic}. As a whole, these experimental advances show that non-Abelian hopping phases are no longer merely theoretical constructs but can be routinely synthesized in state-of-the-art photonic architectures, thereby providing a realistic experimental platform for implementing the non-Abelian gauge couplings considered in the present work.

\subsection{Controllable NN and NNN nonreciprocal couplings}
The proposed tight-binding lattice Hamiltonian is well within the capabilities of current integrated photonic platforms. In evanescently coupled optical waveguide arrays, nearest-neighbor (NN) coupling naturally arises from the overlap of adjacent guided modes and has been extensively demonstrated in femtosecond-laser-written and semiconductor photonic lattices, establishing discrete photonic lattices as versatile realizations of tight-binding Hamiltonians \cite{christodoulides2003discretizing,szameit2010discrete}. More recently, advances in waveguide design and fabrication have enabled precise engineering of next-nearest-neighbor (NNN) coupling through geometric optimization of the lattice, allowing both the strength and even the sign of the second-order hopping amplitude to be experimentally controlled \cite{schulz2022geometric}. In parallel, non-Hermitian photonic lattices incorporating controlled gain, loss, and asymmetric couplings have been experimentally realized on a variety of optical platforms, including PT symmetric waveguide arrays and topological photonic lattices \cite{zeuner2015observation,weimann2017topologically}. Electric circuits have been used in experiments to observe complex energy braiding in non-Abelian non-Hermitian systems \cite{chen2026implementing}. These experimental advances strongly suggest that the simultaneous implementation of NN and NNN couplings in non-Hermitian photonic lattices can be achieved with current fabrication technologies and experimentally feasible integrated photonic settings.

\subsection{Onsite gain and loss}
The non-Hermitian gain and loss elements incorporated in the current model are experimentally realizable using several widely known photonic platforms \cite{el2018non,feng2017non,ozdemir2019parity,miri2019exceptional, gupta2020parity,li2023exceptional}. In optical waveguide arrays, non-Hermiticity can be introduced through spatially engineered optical absorption, selective waveguide pumping, or controlled radiation leakage, thereby enabling precise control over site-dependent gain and loss. The first experimental demonstrations of PT symmetric optical structures employed balanced gain and loss in coupled optical waveguides to observe PT-symmetry breaking, exceptional points, and non-reciprocal light transport \cite{guo2009observation,ruter2010observation, peng2016chiral}. Subsequently, non-Hermitian topological phenomena were experimentally realized in waveguide lattices through controlled dissipation, including the observation of topological phase transitions induced by loss engineering \cite{zeuner2015observation} and the topological edge states in PT-symmetric photonic crystals \cite{weimann2017topologically}, non-Hermitian skin effects in actively controlled photonic lattices \cite{weidemann2020topological}, and EP-enhanced light dynamics in integrated photonic structures \cite{hodaei2017enhanced,chen2017exceptional}. As a result, the gain-loss non-Hermitian terms in the model Hamiltonian are fully compatible with existing experimental capabilities and can be realized using current integrated photonic media such as waveguide arrays and coupled resonator systems.

In this connection, it could be worthwhile to note the experimental demonstrations of non-Abelian effects and/or non-Hermitian skin effects in various closely related settings, including, synthetic frequency lattice \cite{wang2021topological}, time-multiplexed open photonic lattice \cite{parto2023non}, electric circuits \cite{chen2026implementing}, acoustic resonators \cite{zhang2025experimental}, and magnonics \cite{rao2024braiding}, among others. All these experimental advances demonstrate that synthetic non-Abelian SU(2) gauge fields, site-resolved gain and loss, and nonreciprocal NN and NNN couplings are experimentally feasible and can be expected to be implemented in state-of-the-art integrated photonic and other related systems. This could be particularly the case since, despite multiple non-Hermitian degrees of freedom, the system configuration is relatively a simplified version of a far more generic class of complex non-Hermitian systems, and its EP-mediated topological phase transition is analytically tractable.

\bibliography{NH_Braids_bib}
\end{document}